\documentclass[preprint,12pt]{elsarticle}

\usepackage[utf8]{inputenc}
\usepackage{graphicx}
\usepackage{color}
\usepackage{amsmath}
\usepackage{mathtools}
\usepackage{epstopdf}
\usepackage[justification=centering]{caption}
\usepackage[english]{nomencl}

\usepackage{ifthen}
\usepackage{ulem}
\usepackage{subfig}
\usepackage{float}
\usepackage{array}
\usepackage{rotating}
\usepackage{multirow}
\usepackage{multicol}
\usepackage{geometry}
\usepackage{etoolbox}
\usepackage{amssymb} 

\usepackage{xspace}
\usepackage{textcomp, gensymb}
\usepackage{blindtext}
\usepackage{adjustbox}
\usepackage{makecell}

\usepackage{tikz}
\usetikzlibrary{tikzmark, calc}

\usepackage[colorlinks]{hyperref}
\usepackage{lipsum}
\DeclareGraphicsExtensions{.png}

\usepackage{tikz}
\usetikzlibrary{calc}
\usetikzlibrary{math}
\usepackage{pgfplots}
\pgfplotsset{compat=1.14}
\usetikzlibrary{intersections}

\newcommand{\npar}{\vspace{\baselineskip}}

\newcommand{\derive}[2]{\frac{\mathrm{d}{#1}}{\mathrm{d} {#2}}}

\newcommand{\parth}[1]{\left( {#1} \right)}
\newcommand{\crocht}[1]{\left[ {#1} \right]}
\newcommand{\bars}[1]{\left| #1 \right|}
\newcommand{\up}[1]{$^\mathrm{#1}$}

\newcommand{\debm}{\mathrm{kg}/\mathrm{m}^{2}/\mathrm{s}}

\newcommand{\Ja}{\mathrm{Ja}}
\renewcommand{\Re}{\mathrm{Re}}
\renewcommand{\Pr}{\mathrm{Pr}}
\newcommand{\Nu}{\mathrm{Nu}}

\newcommand{\dtheta}{\mathrm{d}\theta}

\renewcommand{\cos}[1]{\mathrm{cos}\parth{#1}}
\renewcommand{\sin}[1]{\mathrm{sin}\parth{#1}}

\renewcommand{\ln}[1]{\mathrm{ln}\parth{#1}}

\newcommand{\etal}{{et al.}\xspace}
\newcommand{\ie}{{i.e.}\xspace}
\newcommand{\eg}{{e.g.}\xspace}

\makenomenclature

\renewcommand*{\nompreamble}{\begin{multicols}{2}}
\renewcommand*{\nompostamble}{\end{multicols}}
\journal{International Journal of Heat and Mass Transfer}

\renewcommand\nomgroup[1]{%
  \item[\bfseries
  \ifstrequal{#1}{L}{Latin symbols}{%
  \ifstrequal{#1}{G}{Greek symbols}{%
  \ifstrequal{#1}{S}{Subscripts}{
  \ifstrequal{#1}{A}{Acronyms}{
  \ifstrequal{#1}{N}{Non-dimensional numbers}{}}}}}%
]}

\newcommand{\blue}[1]{\textcolor{black}{#1}}
\newcommand{\green}[1]{\textcolor{black!50!black}{#1}}
\newcommand{\greenbis}[1]{\textcolor{black!50!black}{#1}}
\newcommand{\bluebis}[1]{\textcolor{black}{#1}}
\newcommand{\greenbbis}[1]{\textcolor{green!50!black}{#1}}

\begin{document}

%Latin symbols
\nomenclature[latin]{$A$}{Wall area proportion [-]}
\nomenclature[latin]{$h$}{Heat transfer coefficient [W/m\up{2}/K]}
\nomenclature[latin]{$T$}{Temperature [K]}
\nomenclature[latin]{$t$}{Time [s]}
\nomenclature[latin]{$N$}{Nucleation sites or bubble density [m\up{-2}]}
\nomenclature[latin]{$R$}{Bubble radius [m]}
\nomenclature[latin]{$D$}{Bubble diameter [m]}
\nomenclature[latin]{$h_{LV}$}{Latent heat of vaporization [J/kg]}
\nomenclature[latin]{$f$}{\green{Bubble departure frequency} [s\up{-1}]}
\nomenclature[latin]{$F_A$}{Bubble influence area enhancement factor [-]}
\nomenclature[latin]{$C_f$}{Friction coefficient [-]}
\nomenclature[latin]{$C_D$}{Drag coefficient [-]}
\nomenclature[latin]{$C_{AM}$}{Added mass coefficient [-]}
\nomenclature[latin]{$g$}{Gravity acceleration [m\up{2}/s]}
\nomenclature[latin]{$U$}{Velocity [m/s]}
\nomenclature[latin]{$K$}{Bubble growth constant [-]}
\nomenclature[latin]{$U_{\tau}$}{Wall friction velocity [m/s]}
\nomenclature[latin]{$L_{c}=\sqrt{\frac{\sigma}{\parth{\rho_{L}-\rho_{V}}g}}$}{Capillary length [m]}
\nomenclature[latin]{$D_{h}$}{Hydraulic diameter [m]}
\nomenclature[latin]{$P$}{Pressure [bar]}
\nomenclature[latin]{$G$}{Mass flux [$\debm$]}
\nomenclature[latin]{$R_{c}$}{Cavity radius [m]}
\nomenclature[latin]{$l_{sl}$}{Bubble sliding length [m]}
\nomenclature[latin]{$s_{a}$}{Average distance between active sites[m]}
\nomenclature[latin]{$s_{b}$}{Average distance between nucleating bubbles [m]}

% \nomenclature[latin]{$r_{w}$}{Bubble foot radius [m]}
% \nomenclature[latin]{$F$}{Force [N]}
% \nomenclature[latin]{$V$}{Volume [m\up{3}]}
% \nomenclature[latin]{$C$}{Force coefficient [-]}
% \nomenclature[latin]{$U_{rel}=U_{L}-U_{b}$}{Relative velocity [m/s]}
% \nomenclature[latin]{$E$}{Kinetic energy [J]}

%Non-Dimensional numbers
\nomenclature[ND]{$\Re_{D_{h}}=\frac{G_{L}D_{h}}{\mu_{L}}$}{Liquid bulk Reynolds number [-]}
\nomenclature[ND]{$\Nu = \frac{hL}{\lambda}$}{Nusselt number [-]}
\nomenclature[ND]{$\Pr =\frac{\nu}{\eta}$}{Prandtl number [-]}
\nomenclature[ND]{$\Ja =\frac{\rho_{L}c_{P,L}\bars{T-T_{sat}}}{\rho_{V}h_{LV}}$}{Jakob number [-]}
\nomenclature[ND]{$y^{+} =\frac{yU_{\tau}}{\nu_{L}}$}{Non-dimensional wall distance [-]}

% \nomenclature[ND]{$\Re_{b}=\frac{U_{rel}D_{b}}{\nu_{L}}$}{Bubble Reynolds number [-]}
% \nomenclature[ND]{$\Sr =\frac{2 \gamma R}{\bars{U_{rel}}}$}{Non-dimensional shear rate [-]}
% \nomenclature[ND]{$\Fr =\frac{\rho_{L}U_{L}^{2}}{\parth{\rho_{L}-\rho_{V}}gR}$}{Froude number [-]}
% \nomenclature[ND]{$\Eo =\frac{\parth{\rho_{L}-\rho_{V}}gR^{2}}{\sigma}$}{Eotvos number [-]}
% \nomenclature[ND]{$\Ca =\frac{\mu_{L}U_{L}}{\sigma}$}{Capillary number [-]}

%Greek symbols
\nomenclature[grec]{$\phi$}{Heat flux [J/m\up{2}/s]}
\nomenclature[grec]{$\lambda$}{Thermal conductivity [W/m/K]}
\nomenclature[grec]{$\eta$}{Thermal diffusivity[m\up{2}/s]}
\nomenclature[grec]{$\rho$}{Density [kg/m\up{3}]}
\nomenclature[grec]{$\sigma$}{Surface tension [J/m\up{2}]}
\nomenclature[grec]{$\theta$, $\dtheta$}{Contact angle and half-hysteresis [$\degree$ or rad]}
\nomenclature[grec]{$\tau$}{Shear stress [kg/m/s\up{2}]}
\nomenclature[grec]{$\nu$}{Kinematic viscosity [m\up{2}/s]}
\nomenclature[grec]{$\mu$}{Dynamic viscosity [J.s/m\up{-3}]}

%Subscripts
\nomenclature[subscript]{$w$}{Wall or wait}
\nomenclature[subscript]{$L$}{Liquid}
\nomenclature[subscript]{$V$}{Vapor}
\nomenclature[subscript]{$c,L$}{Liquid forced convection}
\nomenclature[subscript]{$q$}{Quenching}
\nomenclature[subscript]{$e$}{Evaporation}
\nomenclature[subscript]{$b$}{Bubble}
\nomenclature[subscript]{$coal,st$}{Static coalescence}
\nomenclature[subscript]{$sl$}{Sliding}
\nomenclature[subscript]{$coal, sl$}{Sliding coalesence}
\nomenclature[subscript]{$c,V$}{Vapor forced convection}
\nomenclature[subscript]{$lo$}{Lift-off}
\nomenclature[subscript]{$d$}{Departure}
\nomenclature[subscript]{$g$}{Growth}
\nomenclature[subscript]{$sit$}{Nucleation sites}
\nomenclature[subscript]{$sit,a$}{Active nucleation sites}
\nomenclature[subscript]{$g$}{Growth}
\nomenclature[subscript]{$sat$}{Saturation}

%Acronyms
\nomenclature[acronym]{CHF}{Critical Heat Flux}
\nomenclature[acronym]{PWR}{Pressurized Water Reactor}
\nomenclature[acronym]{SMR}{Small Modular Reactor}
\nomenclature[acronym]{BWR}{Boiling Water Reactor}
\nomenclature[acronym]{CFD}{Computational Fluid Dynamics}
\nomenclature[acronym]{CMFD}{Computational Multi-Fluid Dynamics}
\nomenclature[acronym]{HFP}{Heat Flux Partitioning}
\nomenclature[acronym]{NSD}{Nucleation Site Density}
\nomenclature[acronym]{ITO}{Indium Tin Oxide}

\begin{frontmatter}

\title{On physical considerations regarding development and validation of Heat Flux Partitioning models: application to vertical boiling flows simulations}

\author{Luc Favre \fnref{label1,label2}\corref{cor3}}
\author{Catherine Colin \fnref{label1}}
\author{Stéphane Pujet\fnref{label2}}
\author{Stéphane Mimouni\fnref{label2}}
\fntext[label1]{Institut de M\'ecanique des Fluides de Toulouse (IMFT), Universit\'e de Toulouse, INPT, CNRS, All\'ee Camille Soula, Toulouse, 31400, France}
\fntext[label2]{\'Electricit\'e de France Recherche \& D\'eveloppement (EDF R\&D), 6 Quai Watier, Chatou, 78400,  France}
\cortext[cor3]{First corresponding author, favre.luc05@gmail.com}

\begin{abstract}
\bluebis{This work aims at conducting a critical assessment of wall boiling modeling through the Heat Flux Partitioning approach}. To do so, a new model dedicated to vertical boiling flows is \bluebis{constructed}, with a revisited partitioning including \green{an evaporation heat flux} related to bubble coalescence \bluebis{while discussing and assessing each modeling step}. 

Closure laws include a recent model for bubble dynamics, a new correlation for the bubble maximum lift-off diameter, and comprehensive selection of existing models for nucleation site density and bubble wait time.

Each formulation is compared to relevant existing data from the literature in order to emphasize the importance of separate validation in such a modeling framework.

The whole model is then confronted to detailed wall boiling experiments to simultaneously compare boiling curve predictions along other physical parameters such as boiling time scales (bubble growth, transient conduction, bubble wait), \green{bubble departure frequency}, or nucleation site density.

Finally, validation against wall temperature measurements in various conditions are used to assess the model accuracy and further discuss the limits of the Heat Flux Partitioning approach.

\end{abstract}


\end{frontmatter}

\printnomenclature

\section{Introduction}\label{sec:intro}

Boiling phase change is one of the most efficient physical phenomenon to operate heat transfer. However, when one reaches the Critical Heat Flux (CHF), the phase change departs from nucleate boiling and instantly switch to film boiling, resulting in a large decrease of the global heat transfer coefficient and posing overheating risks to the heated material. Boiling is thus a key element of many industrial processes involving heat transfer, e.g. nuclear energy (\blue{Pressurized Water Reactors, Small Modular Reactors, Boiling Water Reactors}), space industry (electronic devices cooling, cryogenic tanks safety) or various power cycles (heat pumps, air conditioning, etc.).

Studying the underlying physics behind boiling has been an internationally active research field for decades. In particular, the evolution of Computational Multi-Fluid Dynamics (CMFD) applications has created a need for dedicated models to represent wall boiling within simulations in order to respectively evaluate the portion of heat flux received by the liquid and that contributing to vapor generation. 

At first, direct correlations were used to achieve this goal \cite{jens_analysis_1951, thom_boiling_1967}. Now, several CMFD codes rely on a Heat Flux Partitioning (HFP) model \cite{guelfi_neptune_2007, favre_neptune_2022},which goal is to split the total applied heat flux between different heat transfer mechanisms. The most notable model of this kind has been initially proposed by Kurul \& Podowski \cite{kurul_multidimensional_1990} who divided the heat flux between convection towards the liquid $\phi_{c,l}$, boiling $\phi_{b}$ and transient conduction induced by bubble departure (quenching, $\phi_{q}$) (Eq. \ref{eq:hfp_KP}):

\begin{equation}
    \phi_{w} = \phi_{c,L} + \phi_{e} + \phi_{q}
    \label{eq:hfp_KP}
\end{equation}

\begin{equation}
    \phi_{c,L} = A_{c,L} h_{c,L} \parth{T_{w} - T_{L}}
    \label{eq:phicl_KP}
\end{equation}

\begin{equation}
    \phi_{q} = A_{q} \dfrac{2 \lambda_L \parth{T_w - T_L}}{\sqrt{\pi \eta_L t_q}}
    \label{eq:phiq_KP}
\end{equation}

\begin{equation}
    \phi_{e} = N_{sit} f \dfrac{4}{3} \pi R_{b}^3 \rho_{v} h_{LV} 
    \label{eq:phib_KP}
\end{equation}

While developing the expressions of these three fluxes (Eqs. \ref{eq:phicl_KP}, \ref{eq:phiq_KP} and \ref{eq:phib_KP}), Kurul \& Podowski had to select dedicated models to compute key physical parameters in the heat flux partitioning, namely:

\begin{itemize}
    \item The liquid turbulent heat transfer coefficient $h_{c,L}$, expressed using the Stanton number in the liquid buffer layer ;
    \item The nucleation site density $N_{sit}$ which was computed using the correlation of Lemmert \& Chawla \cite{lemmert_influence_1977} ;
    \item The \green{bubble departure frequency} $f$, supposed equal to the inverse of the bubble wait time $1/t_w$ and following the relationship of Cole \cite{cole_bubble_1967} ;
    \item The bubble lift-off diameter $D_b = 2 R_b$, relying on the formulation of \"Unal \cite{unal_maximum_1976} for high pressure flows. \green{Though \"Unal refers to this diameter as "departure" diameter, we prefer using the term "lift-off" in the frame of this work as the diameter at which the bubble leaves the surface.}
\end{itemize}

The areas associated to liquid and quenching heat transfer are computed as:

\begin{equation}
    A_{q} = 1-A_{c,L}=\mathrm{min}\parth{1\ ;\ F_A N_{sit} \pi R_b^2}
\end{equation}
where $F_A=4$ accounts for the bubble area of influence when leaving the surface.

Their model was finally validated against vertical flow boiling temperature measurements.

\npar

Later, this modeling approach has been pursued by various authors proposing different upgrades to the heat flux partitioning to further account for detailed physical phenomena involved in wall boiling. Research works on HFP models generally involve extended expressions for heat fluxes (\eg to include the effect of bubble sliding) or revisited closure laws for specific physical parameters.

For instance, Basu \etal \cite{basu_wall_2005} accounted for bubble sliding in vertical flow boiling and relied on experimental results (\eg from Maity \cite{maity_effect_2000}) to propose a full set of new closures laws which was validated for low pressure and liquid mass flux cases. In 2017, Gilman \& Baglietto \cite{gilman_self-consistent_2017} implemented a mechanistic force balance in the HFP model to compute the bubble departure and lift-off diameter while considering a solid quenching term in the partitioning along with a novel nucleation site suppression mechanism. They validated their model against low pressure vertical boiling experiments at large liquid mass fluxes. Among the latest models, we can mention that of Zhou \etal \cite{zhou_mechanistic_2021} who included the effect of condensation at bubble top in their HFP model and used  correlations for bubble departure and lift-off diameter developed using on their own experimental results \cite{zhou_experimental_2020}. 

\npar

Validation of HFP models is often achieved by comparing wall temperature predictions with experimental results \cite{gilman_self-consistent_2017, zhou_mechanistic_2021} or through bulk void fraction profile comparisons in CFD simulations using the considered HFP model \cite{wang_development_2023, li_development_2018}. However, due to the very large number of physical variables at stake, such validation process is quite incomplete regarding the variety of involved physical sub-parameters. As a simple example, a 50\% underestimation in the nucleation site density combined with an overestimation of 100\% in the \green{bubble departure frequency} would result in the exact same evaporation heat flux (Eq. \ref{eq:phib_KP}) while failing to properly capture those boiling features. Moreover, such models often include empirical parameters (\eg contact angle or bubble growth constant) and optimizing their values to match total heat flux measurements will hardly ensure proper validation if extensive comparisons with detailed boiling measurements (\eg bubble diameter, nucleation site density, etc.) is not achieved beforehand. Therefore, missing a proper separate validation of the different closure laws is very likely to result in mutually compensating errors, though the output of the whole model regarding void fractions and wall temperatures could be accurate.

On the other hand, a large number of detailed boiling experiments have been conducted over the last years which have given access to detailed measurements of key parameters for HFP models. For instance, Maity \cite{maity_effect_2000} observed detailed bubble dynamics and contact angle in various heater orientation  \cite{maity_effect_2000}, Situ \cite{situ_bubble_2005} has measured several bubble lift-off diameters in vertical flow boiling, Estrada-Perez \etal \cite{estrada-perez_time-resolved_2018} quantified the thermal impact of bubble sliding in vertical boiling, Richenderfer \cite{richenderfer_investigation_2018} and Kossolapov \cite{kossolapov_experimental_2021} have developed novel experimental apparatus to measure the heat flux distribution and bubble dynamics in vertical flow boiling at low and high pressure respectively. This non-exhaustive list of novel and insightful experimental results should now be a motivation to perform finer validation of the various sub-models involved in the construction of such wall boiling formulations \cite{favre_modelisation_2023}. \blue{Moreover, such experiments \cite{richenderfer_experimental_2018, zupancic_wall_2022, kossolapov_experimental_2021} have also shown that the concept of the heat flux partitioning is actually a proper way to decompose the total heat transfer from the wall to the fluid by reconstructing the experimental applied heat flux through recalculation of individual components such as liquid convection, evaporation and quenching.}

\bluebis{In this paper, we propose to discuss the physical consistency of this approach} by developing a Heat Flux Partitioning model dedicated to vertical flow boiling that includes bubble sliding and coalescence, to perform a detailed validation of the selected closure laws while ending up with a reduced set of remaining empirical parameters. \bluebis{The goal here is not to claim that a best-performing model has been developed, but rather to discuss traditional hypotheses made in the frame of HFP modeling and explore ways to ensure physical consistency.} In Section \ref{sec:hfp_model}, we describe the proposed HFP formulation before first focusing on single phase heat transfer and bubble dynamics in Section \ref{sec:bub_dyn}. Then focus is put on the bubble nucleation cycle and time scales calculation in Section \ref{sec:bubble_times}, followed by the modeling of nucleation site density and bubble interactions in \ref{sec:bubble_interactions}. Finally, Section \ref{sec:HT_areas} details the remaining closure laws before focusing in Section \ref{sec:validation} on the whole model validation first with a detailed experimental case at 10.5 bar from Kossolapov \cite{kossolapov_experimental_2021} followed by heat flux / wall temperature predictions for a broader range of experimental conditions and discussing the overall model development.

\section{Heat Flux Partitioning Description}
\label{sec:hfp_model}
\subsection{Chosen Partitioning}

The concept of Heat Flux Partitioning is to compute the proportion of the total applied wall heat flux that is directly received by the liquid phase or that contributes to vapor generation. In reality, a large majority of the heat flux is first transferred to the liquid phase before contributing to vapor generation through evaporation of the superheated liquid near the liquid-vapor interface \cite{kossolapov_experimental_2021}. Regarding CFD applications, the interest of using HFP models is to estimate the final portion of the wall heat flux that is used to heat the liquid and to grow vapor bubbles respectively. Therefore, in this work, the considered \green{evaporation heat flux} corresponds to the final volume of vapor generated at the wall before bubbles leave the surface towards the bulk flow. \blue{In addition, we also define the bubble departure as the moment when the bubble leaves its nucleation site by sliding and bubble lift-off when the bubble detaches from the boiling surface.}

\npar

\begin{figure}[H]
\centering
\includegraphics[width=1.0\linewidth]{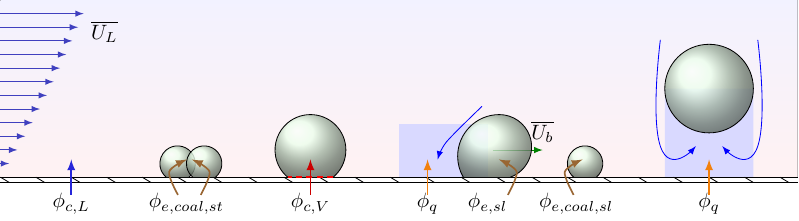}
\caption{Sketch of the HFP model.}
\label{fig:hfp_model}
\end{figure}

In vertical flow boiling, bubble sliding is nearly always observed before lift-off \cite{kossolapov_experimental_2021, maity_effect_2000, estrada-perez_time-resolved_2018, situ_bubble_2005}. Moreover, at high pressure and wall superheat, the nucleation site density becomes very large, making wall bubble coalescence highly probable and can often be identified as a lift-off trigger mechanism \cite{scheiff_experimental_2021, prodanovic_bubble_2002, thorncroft_experimental_1998}. Finally, when approaching the Critical Heat Flux (CHF), the wall dry area in direct contact with vapor increases significantly. In the proposed Heat Flux Partitioning model we wish to account for those aforementioned phenomena, which is why the total heat flux is proposed to be split between the following different heat transfer mechanisms (Figure \ref{fig:hfp_model}):

\begin{enumerate}
    \item A forced convection heat flux towards the liquid phase $\phi_{c,L}$ ;
    \item A quenching transient heat flux towards the liquid resulting from bubble sliding and lift-off $\phi_{q}$ ;
    \item An \green{evaporation heat flux} $\phi_{e}$ divided in three types of bubbles dynamics. A first one associated to static bubble coalescence when they are close enough to merge and lift-off before starting to slide ($\phi_{e,coal,st}$), a second one related to sliding growing bubbles ($\phi_{e,sl}$) and a last one corresponding to the coalescence of static growing bubbles on their nucleation site with sliding bubbles ($\phi_{e,coal,sl}$) ;
    \item A conductive heat flux towards the vapor phase through the wall dry areas $\phi_{c,V}$.

\end{enumerate}

\blue{This partitioning of the total heat flux is not dependent on the heater orientation and could very well be applied both for vertical and horizontal boiling. However, the closure laws that will be involved in the calculation of the different fluxes can be gravity orientation dependent and thus be developed and validated only for vertical boiling flows.}

\subsection{Liquid forced convection heat flux}

The forced convection towards the liquid is classically expressed as :

\begin{equation}
    \phi_{c,L} = A_{c,L} h_{c,L} \parth{T_{w} - T_{L}}
    \label{eq:phi_cL}
\end{equation}

where we will require closure relationships for the forced convection area $A_{c,L}$ and heat transfer coefficient $h_{c,L}$.

\subsection{Quenching heat flux}

Assuming that the quenching resulting from cold liquid displacement in the bubble wake leads to a transient heat transfer analogous to heat conduction in an infinite liquid medium \cite{del_valle_subcooled_1985}, we can write:

\begin{equation}
    \phi_{q} = A_{q} \dfrac{2 \lambda_{L} \parth{T_w - T_L}}{\sqrt{\pi \eta_L t_q}}
    \label{eq:phi_q}
\end{equation}

where we require closure laws for the quenching area $A_q$ and quenching time $t_q$.

Note that the temperature difference between the wall and bulk liquid can be rewritten as $T_w - T_L = \Delta T_w + \Delta T_L$ where $\Delta T_w = T_w - T_{sat}$ is the wall superheat and $\Delta T_L = T_{sat} - T_L$ is the liquid subcooling, which are useful variables that will be used later in this work.

\subsection{Evaporation heat flux}

The total \green{evaporation heat flux} writes:

\begin{equation}
    \phi_{e} = \phi_{e,coal,st} + \phi_{e, sl} + \phi_{e,coal,sl}
\end{equation}

If bubbles coalesce while still being attached to their nucleation site, their radius can not be larger than the departure by sliding radius $R_d$. In this model, we assume that static coalescence would occur when they are close to $R_d$, yielding:

\begin{equation}
    \phi_{e,coal,st} = N_{sit,coal,st} f \rho_V h_{LV} \dfrac{4}{3}\pi R_{d}^3
    \label{eq:phie_st}
\end{equation}

where we need closure laws for the nucleation sites leading to static coalescence bubbles $N_{sit,coal,st}$, the \green{bubble departure frequency} $f$ and the bubble departure diameter $R_d$.

Following the previous assumption, we suppose that a static bubble on its site coalescing with an upcoming sliding bubble will have a radius close to $R_d$ while the sliding bubble has a larger sliding radius $R_{sl}$. Reasoning in terms of pairs of bubbles (sliding and static) yields the following two \green{evaporation fluxes}:

\begin{equation}
    \phi_{e,sl} = N_{sit,sl} f \rho_V h_{LV} \dfrac{4}{3}\pi R_{sl}^3
    \label{eq:phie_sl}
\end{equation}

\begin{equation}
    \phi_{e,coal,sl} = N_{sit,coal,sl} f \rho_V h_{LV} \dfrac{4}{3}\pi R_{d}^3
    \label{eq:phie_coal_sl}
\end{equation}

where we need closure laws for the nucleation sites leading to sliding and coalesced bubbles $N_{sit,sl}$ and $N_{sit,coal,sl}$ along with the sliding bubble diameter $R_{sl}$.

\subsection{Vapor conductive heat flux}

In wall boiling regime, the vapor in direct contact with the heater remains mostly quiescent within the bubbles. Therefore, we can expect the direct heat transfer between the wall and the vapor phase to be of conductive nature. The typical length over which this heat conduction operates is then the average bubble radius $\left<R\right>$. Assuming the vapor to be at saturation temperature, we thus have:

\begin{equation}
    \phi_{c,V} = A_{c,V} \dfrac{\lambda_V}{\left<R\right>} \parth{T_{w} - T_{sat}}
\end{equation}

where closure are required for $A_{c,V}$ and $\left<R\right>$.

\section{Single-phase heat transfer and bubble dynamics}
\label{sec:bub_dyn}

\subsection{Single-phase heat transfer}

The first closure relationship required for the Heat Flux Partitioning is the liquid forced convective heat transfer coefficient $h_{c,L}$. Though it is very commonly used, it has to be validated against experimental measurements for the aimed application of the model. As stated in the introduction, we focus on vertical flow boiling cases for which we have gathered single-phase measurements from four upwards boiling flow experiments (Table \ref{tab:exp_data_convection}). It covers a large range of operating conditions in terms of pressure, liquid mass flux, subcooling and wall heat flux.

\begin{table}[h!]

%\begin{changemargin}{-1cm}{0cm}

\noindent\makebox[\textwidth]{

\scriptsize
\centering
\begin{tabular}{p{27mm}|c c c c c c c c} 
Author & $D_{h}$ [mm] & $P$ [bar] & $G_{L}$ [$\debm$] & $\Delta T_{L}$ [K] & $\phi_{w}$ [MW/m\up{2}] & $T_{sat}-T_{w}$ [K] & $N_{mes}$ [-] \\
\hline
\\

Kossolapov \cite{kossolapov_experimental_2021} \newline (2021) & 12 & 10.5 & 500 - 2000 & 10 & 0.1 - 0.6  & 0.22 - 9.5 & 12 \\

Richenderfer \cite{richenderfer_experimental_2018} \newline (2018) & 15 & 1 - 5 & 1000 - 2000 & 10-20 & 0.1 - 0.63 & 1 - 18.7 & 13 \\

Jens-Lottes \cite{jens_analysis_1951} \newline (1951) & 5.74 & 137.9 & 2617.5 & 53.3 - 92.2 & 0.91 - 2.37 & 0.33 - 44.1 & 15 \\

Kennel \cite{kennel_local_1949} \newline (1948) & 4.3 - 13.2 & 2 - 6.2 & 284 - 10~577 & 11.1 - 83.3 & 0.035 - 1.89 & 0.35 - 69 & 52 \\
\hline
\end{tabular}
}
\caption{Experimental data range of wall temperature measurements from the single-phase part of boiling curves. $N_{mes}$ is the number of measurements of each data set.}
\label{tab:exp_data_convection}
\end{table}

\begin{figure}[H]
\centering
\includegraphics[width=0.7\linewidth]{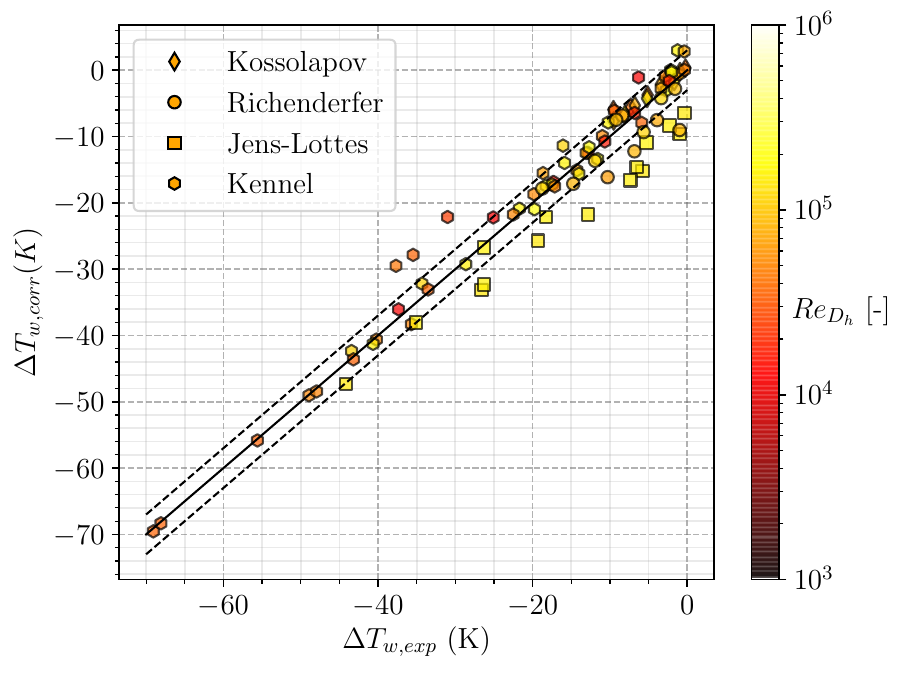}
\caption{Single-phase wall temperature predictions using Gnielinski correlation with $\pm 3$K error bars. Colorbar represents the bulk flow Reynolds number.}
\label{fig:hcL_gnielinski}
\end{figure}

After testing several correlations, it was found that the formulation of Gnielinski \cite{gnielinski_new_1975} (Eq. \ref{eq:gnielinski}) is able to reproduce most of the wall temperature measurements from the considered experiments with an average error usually smaller than 3K. Global comparison with the selected experiments is presented on Figure \ref{fig:hcL_gnielinski}).

\begin{equation}
\Nu = \frac{\dfrac{C_{f}}{2} \parth{\Re_{D_h} - 1000} \Pr }{1+12.7\sqrt{\dfrac{C_{f}}{2}} \parth{\Pr^{2/3}-1}}
\label{eq:gnielinski}
\end{equation}

It must be noted that this correlation will be kept in the two-phase nucleate boiling regime, though some authors suggest that the value of $h_{c,L}$ should be modified when bubbles appear on the surface. For instance, Basu \etal \cite{basu_wall_2005} suggest to enhance the liquid heat transfer coefficient by 30\%. Also, Mimouni \etal \cite{mimouni_second_2010} and Gilman \& Baglietto \cite{gilman_self-consistent_2017} use a modified liquid velocity profile in the wall vicinity accounting for the pseudo-roughness induced by the bubbles, thus impacting the liquid heat transfer coefficient. On the other hand, Kossolapov \cite{kossolapov_experimental_2021} extrapolated the single-phase heat transfer coefficient to the boiling region to estimate the liquid forced convection component in his experiments and Zhou \etal \cite{zhou_mechanistic_2021} considered a single-phase liquid velocity and temperature profile to compute the liquid heat transfer coefficient for both single-phase and two-phase regimes. Therefore, in absence of detailed data allowing the precise assessment of a potential enhancement of $h_{c,L}$ in the boiling region, we choose to compute it using Gnielinski correlation (Eq. \ref{eq:gnielinski}) everywhere.

\subsection{Bubble dynamics}

\subsubsection{Bubble departure diameter}

\begin{figure}[H]
\centering
\includegraphics[width=0.6\linewidth]{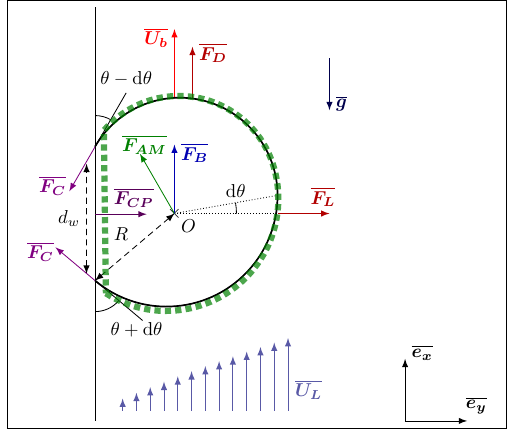}
\caption{Sketch of the force balance model for bubble departure and sliding. Adapted from Favre \etal \cite{favre_updated_2023}.}
\label{fig:bub_bdf}
\end{figure}

Prediction of the bubble departure diameter by sliding is achieved using a mechanistic force balance on a single bubble as depicted on Figure \ref{fig:bub_bdf}. 

The full force balance model is detailed in Favre \etal \cite{favre_updated_2023, favre_analytical_2022} and relies on solving the following equation:

% \begin{equation}
% %C_{AM,x}K^{2} \frac{\Ja_{w}^{2}}{\Pr_{L}} + \frac{1}{3}\frac{\Re_{b}}{\Fr} + \frac{1}{8}C_{D}\Re_{b} = \frac{1}{2} \frac{f_{C,x}}{\Ca}
% \label{eq:pred_nogr}
% \end{equation}

\begin{align}
\label{eq:pred_nogr}
\nonumber - \underbrace{\pi R\sigma f_{C,x}}_{F_{C,x}} + \underbrace{\frac{4}{3}\pi R^{3}\parth{\rho_{L}-\rho_{V}}g}_{F_B} &+ \underbrace{\frac{1}{2}C_{D}\rho_{L}\pi R^{2} U_{L}^{2}}_{F_D} \\
& + \underbrace{\frac{4}{3}\pi R^{3}\rho_{L}~3C_{AM,x}\frac{\dot{R}}{R}U_{L}}_{F_{AM,x}} = 0
\end{align}

where

\begin{equation}
f_{C,x}=2.5\ \frac{\dtheta}{\parth{\pi/2}^{2}-\dtheta^{2}}\sin{\theta}\cos{\dtheta}
\end{equation}

The different terms in Eq. \ref{eq:pred_nogr} represents, by order of appearance: the capillary force, the buoyancy, the drag force and the added mass force. Written this way, it means that bubble departure from its nucleation site is considered when the detaching forces (buoyancy, drag and added mass) overcome the capillary term, where $C_{D}$ is computed following the recent formulation of Shi \etal \cite{shi_drag_2021} and $C_{AM,x} = 0.636$ as detailed in Favre \etal \cite{favre_updated_2023}. Estimating the capillary term remains complicated and requires values of the average contact angle $\theta$ and the contact angle half-hysteresis $\dtheta$ for which no generic values can be set. They both have to be chosen in accordance with the material of the boiling surface and the liquid for $\theta$, and with the operating conditions for $\dtheta$ (\eg low pressure boiling creates larger bubbles that are likely to reach larger tilt, thus bigger $\dtheta$).

By accepting a $5\degree$ uncertainty of $\theta$ and $\dtheta$, this model was reasonably validated for 122 measurements of departing bubbles in vertical boiling with pressures ranging from $1$ to $40$ bar, liquid mass fluxes up to $1654 \debm$, liquid subcooling from $0.3$ to $51$ K and heat fluxes from $2.83$ to $643.2$ kW/m\up{2}.

The bubble growth is modeled as a traditional heat diffusion model, proven valid for early growth stages in subcooled conditions or high pressure boiling with small bubbles \cite{kossolapov_experimental_2021, plesset_growth_1954, klausner_vapor_1993}:

\begin{equation}
R\parth{t} = K\Ja_{w} \sqrt{\eta_{L}t}
\end{equation}
where $K$ is an adjustable constant around the unity. In Favre \etal \cite{favre_updated_2023}, using a value of approximately 0.27 allowed an acceptable predictability for the considered database. However, this value shall be adjusted depending on the flow conditions (\eg decreasing with liquid subcooling and velocity, or increasing with wall heat flux).%, often expressed as $K=\frac{2b}{\sqrt{\pi}}$. In Favre \etal \cite{favre_updated_2023}, Yoo \etal \cite{yoo_development_2018} value of $b=0.24$ was selected since it allowed to match the considered experimental measurements.

\green{In Eq. \ref{eq:pred_nogr}, an estimation of the liquid velocity $U_L$ is required, for which we use Reichardt formula \cite{reichardt_vollstandige_1951} to obtain the local liquid velocity in the near wall region at the bubble center, \ie applied for $y=R$:}

\begin{align}
\dfrac{U_L}{U_{\tau}} = U_{L}^{+} =& \frac{1}{\kappa}\ln{1+\kappa y^{+}} + c \parth{1-e^{-y^{+}/\chi} + \frac{y^{+}}{\chi}e^{-y^{+}/3} }\\
\end{align}
with $\kappa = 0.41$, $\chi = 11$ and $c=7.8$.

The friction velocity $U_{\tau}$ is computed using Mac Adams correlation \cite{mcadams_heat_1954}.

\begin{align}
U_{\tau} =& \sqrt{\frac{\tau_{w}}{\nu_{L}}}\\
\tau_{w} =& 0.018~ \Re_{D_{h}}^{-0.182}~ \rho_{L}\left<U_{L}\right>^{2}
\end{align}

\subsubsection{Bubble sliding}
\label{sec:sliding}

The sliding phase, which plays a great role in enhancing the quenching heat flux along with being a source of bubble coalescence, is computed through an iterative solving of the following equation coming from the force balance:

\begin{align}
\nonumber \parth{1+\frac{\rho_{L}}{\rho_{V}}C_{AM,x}}\derive{U_{b}}{t} = & \parth{\frac{\rho_{L}}{\rho_{V}}-1}g + \frac{3}{8}\frac{C_{D}}{R}\frac{\rho_{L}}{\rho_{V}}\parth{U_{L}-U_{b}}\bars{U_{L}-U_{b}} \\
&+ 3\frac{\dot{R}}{R}\crocht{C_{AM,x}\frac{\rho_{L}}{\rho_{V}}\parth{U_{L}-U_{b}}-U_{b}} - \frac{3}{4}\frac{\sigma}{\rho_{V}}\frac{f_{C,x}}{R^{2}}
\label{eq:ub_dot}
\end{align}

This equation has already shown its capability to capture bubble sliding velocity profiles both at low and high pressures providing a correct bubble growth rate \cite{favre_updated_2023}. \blue{Moreover, these single bubble force balance approaches proved to be able to reproduce the bubble departure and sliding dynamics even for multiple nearby nucleation events \cite{scheiff_experimental_2021, favre_updated_2023, kossolapov_bubble_2024}, supporting that multiple bubble interactions are somewhat negligible in the average bubble dynamics on the boiling surface.}

\subsection{Bubble lift-off diameter}

The physical mechanisms behind bubble lift-off are very diverse. Several experimental works have observed different phenomena triggering the lift-off of a growing bubble from a vertical surface. For instance:

\begin{itemize}
    \item Single bubble lift-off at atmospheric pressure, as in Situ \etal \cite{situ_bubble_2005} or Maity \cite{maity_effect_2000}
    \item Oscillating motion near the wall after a first lift-off, sometimes leading to bubble re-attachment as in Yoo \etal \cite{yoo_experimental_2016}
    \item Lift-off after coalescence with an other bubble, which has been observed by Scheiff \etal \cite{scheiff_experimental_2021}, Prodanovic \etal \cite{prodanovic_bubble_2002} and Thorncroft \etal \cite{thorncroft_experimental_1998}.
\end{itemize}

In particular, Prodanovic \etal \cite{prodanovic_bubble_2002} concluded that single bubble lift-off could not be interpreted as a generic lift-off mechanism due to the scarcity of such events in their experiment. Therefore, as mentioned in Section \ref{sec:hfp_model}, a more general trigger for bubble lift-off could more likely be associated to coalescence between two bubbles or a strong deformation of the bubble shape as suggested by Okawa \etal \cite{okawa_observation_2018} and discussed in Yoo \etal \cite{yoo_experimental_2016}.

In this work, we will consider two possibilities: 
\begin{itemize}
    \item Two bubbles coalescing will lift-off ;
    \item If coalescence is very unlikely (\eg very low nucleation site density), bubble will lift-off when reaching a given maximum diameter.
\end{itemize}

\blue{
The case when a bubble leaves the wall and then re-attach to it \cite{yoo_experimental_2016} is not accounted for here since its occurrence is hard to anticipate and doesn't seem to be a predominant behavior \cite{kossolapov_experimental_2021}.}

The bubble maximum diameter hypothesis is considered to ensure bubble lift-off in the model, avoiding situations where bubbles would slide over very long distances which would make it nonphysical. To estimate this maximum bubble diameter, we propose to build a dedicated correlation based on a large database of experimental measurements from the literature. Considered experiments are presented on Table \ref{tab:exp_data_dlo}.

\begin{table}[ht]

%\begin{changemargin}{-1cm}{0cm}
%\rowcolors{1}{}{lightbrown}
\noindent\makebox[\textwidth]{

\scriptsize
\centering
\begin{tabular}{p{20mm}|c c c c c c c c} 
Author & Fluid & $D_{h}$ [mm] & $P$ [bar] & $G_{L}$ [$\debm$] & $\Delta T_{L}$ [K] & $\phi_{w}$ [kW/m\up{2}] & $\Delta T_{w}$ [K] & $D_{lo}$ [mm] ($N_{mes}$)\\
\hline
\\
Gunther \cite{gunther_photographic_1951} \newline (1951) & Water & 6.92 & 1 - 1.7 & 1492 - 6070 & 33 - 86 & 2.3 - 10.64 & N.A. & 0.32 - 1.02 (12)\\
\\
Griffith \cite{griffith_void_1958} \newline (1958) & Water & 12.7 & 34.5 - 103 & 4651 - 7593 & 11 - 80 & 3.25 - 8.53 &  N.A. & 0.081 - 0.146 (6) \\
\\
Treshchev \cite{treshchev_number_1969} \newline (1969) & Water & 10.18 & 5 - 50 & 1643 - 1789 & 30 - 62 & 1.4 - 2.9 & N.A. & 0.12 - 0.26 (3) \\
\\
Tolubinsky \cite{tolubinsky_vapour_1970} \newline (1970) & Water & 10 & 1 - 10 & 72.6 - 198.4 & 5 - 60 & 0.47 & N.A. & 0.19 - 1.24 (9)\\
\\
\"Unal \cite{unal_maximum_1976} \newline (1976) & Water & 8 & 139 - 177 & 2082 - 2171 & 3 - 5.9 & 0.38 - 0.55 &  N.A. & 0.11 - 0.18 (7) \\
\\
Maity \cite{maity_effect_2000} \newline (2000) & Water & 20 & 1.01 & 0 - 239.6 & 0.3 - 0.7 & N.A. & 5 - 5.9 & 1.8 - 2.4 (4) \\
\\
Prodanovic \cite{prodanovic_bubble_2002} \newline (2002) & Water & 9.3 & 1.01 - 3 & 76.7 - 815.8 & 10 - 60 & 0.1 - 1.2 & N.A. & 0.366 - 2.68 (44)\\
\\
Situ \cite{situ_bubble_2005} \newline (2005)& Water & 19.1 & 1.01 & 471.8 - 910.8 & 1.5 - 20 & 0.06 - 0.2 & N.A. & 0.145 - 0.605  (90)\\
\\
Chu \cite{chu_bubble_2011} \newline (2010) & Water & 22.25 & 1.45 & 301 - 702 & 3.4 - 22.6 & 0.135 - 0.201 &  N.A. & 0.51 - 1.71 (14) \\
\\
Ahmadi \cite{ahmadi_bubble_2012} \newline (2012) & Water & 13.3 & 0.96 - 1.13 & 169 - 497 & 8.4 -20.6 & 0.16 - 0.318 &  11.4 - 18.4 & 0.12 - 3.9 (13) \\
\\
Okawa \cite{okawa_observation_2018} \newline (2018) & Water & 14 & 1.27 - 1.86 & 252 - 490 & 10 - 39 & 0.161 - 0.487 &  N.A. & 0.64 - 0.188 (10) \\
\\
%Samaroo \cite{samaroo_evaluation_2023} \newline (2023) & Water & 6.35 & 1 - 10 & 589 - 1273 & 3 - 15.5 & 40.9 - 84.5 &  1.4 - 18.1 & 0.64 - 0.188 (10) \\
%\\
\hline
\end{tabular}
}

\caption{Bubble lift-off diameters data sets in vertical flow boiling.}
\label{tab:exp_data_dlo}

%\end{changemargin}

\end{table}

We choose to compute the value of the non-dimensional lift-off diameter $D_{lo} / L_{c}$ using the liquid Prandtl number $\Pr_{L} = \nu_L/\eta_L$ at saturation, the density ratio $\rho^{*} = \rho_{L}/\rho_{V}$ (scaling the pressure), the reduced Jakob numbers $\Ja_{w}^{*} = c_{p,L}\parth{T_w-T_{sat}}/h_{LV}$ and $\Ja_{L}^{*} = c_{p,L}\parth{T_{sat}-T_{L}}/h_{LV}$ and the Reynolds number based on the friction velocity and the capillary length $\Re_{\tau} = \rho_L U_{\tau} L_c / \mu_L $. A multi-linear regression yields: 

\begin{equation}
\frac{D_{lo}}{L_{c}} = e^{8.43} \Pr_{L}^{-0.005} \parth{\frac{\rho_{L}}{\rho_{V}}}^{-0.36} {\Ja_{w}^{*}}^{1.15} \parth{1+\Ja_{L}^{*}}^{-6.68} \parth{1+\Re_{\tau}}^{-0.53}
\label{eq:correl_dlo}
\end{equation}

\begin{figure}[H]
\centering
\includegraphics[width=1.0\linewidth]{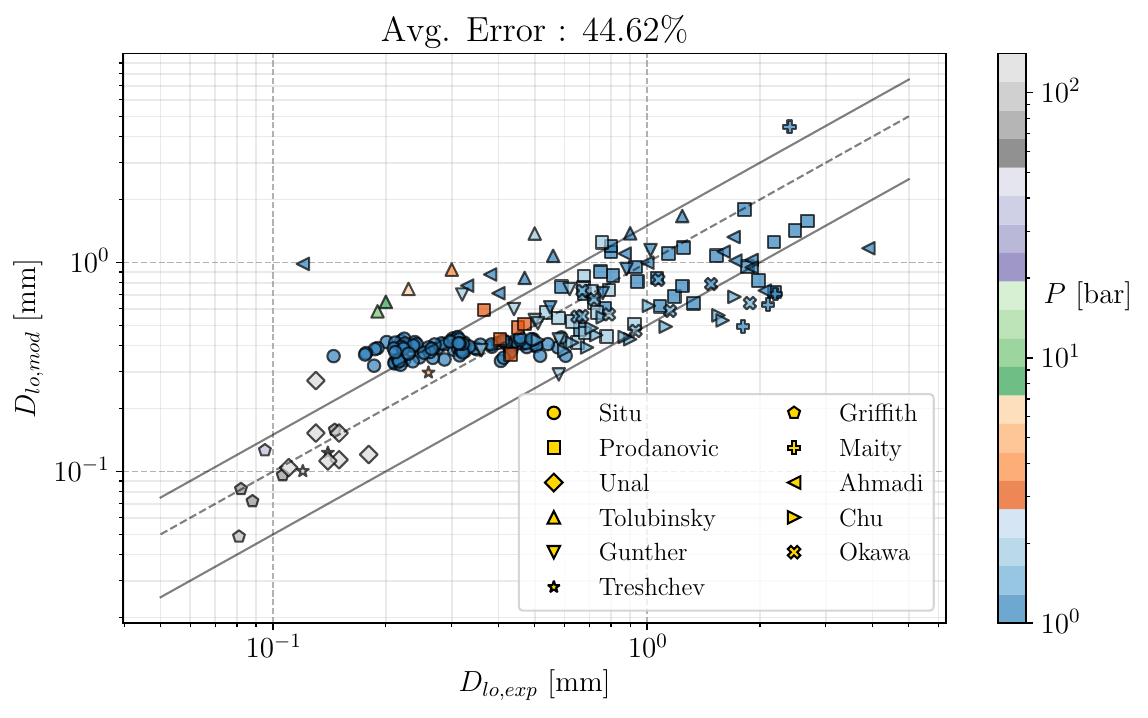}
\caption{Lift-off diameter reduced correlation, $\pm 50\%$ errors bars are indicated.}
\label{fig:lift-off_correl}
\end{figure}

By correlating $\parth{1+\Ja_{L}^{*}}$ and $\parth{1+\Re_{\tau}}$, the formulation degenerates those terms to 1 for saturated and pool boiling conditions. This simple care is often forgotten in similar approaches \cite{kommajosyula_development_2020, zhou_experimental_2020, basu_wall_2005} where the resulting correlations will either diverge or tend to 0 when reaching those conditions. Comparisons of this formulation versus measurements is presented on Figure \ref{fig:lift-off_correl}.

Overall, the proposed correlation produces an average error lower than 50\% on the whole database. Yet being far from perfect, such an accuracy could be considered as acceptable versus other similar correlations usually built for a single given dataset \cite{zhou_experimental_2020, basu_wall_2005} and considering the variety of flow conditions covered by the selected experiments. Moreover, as previously mentioned, the intrinsic goal of this formulation is to provide a somewhat physical-based upper threshold to the bubble diameter while sliding.

\section{Bubble nucleation cycle time scales}
\label{sec:bubble_times}

\subsection{\green{Bubble departure frequency}}

The \green{bubble departure frequency} is a key parameter of the HFP model as it is directly proportional to the boiling and quenching heat fluxes. Considering an established bubble nucleation cycle, we first have a bubble growing on a nucleation site during a growth time $t_{g,d}$ before it departs (by sliding or lift-off) and then a wait time $t_{w}$ occurs during which the thermal boundary layer re-develops following quenching induced by bubble movement, whence a new nucleation event occurs and restarts the whole cycle.

In this context, the \green{bubble departure frequency} is naturally computed as:

\begin{equation}
    f = \dfrac{1}{t_{g,d} + t_{w}}
\end{equation}

One of the most common formulation to compute the \green{bubble departure frequency} has been proposed by Cole \cite{cole_bubble_1967}, who derived the following expression by computing a force balance between buoyancy and surface tension in horizontal pool boiling:

\begin{equation}
    f = \sqrt{\dfrac{4}{3} \dfrac{g \parth{\rho_V - \rho_L}}{\rho_L D_d}}
\end{equation}

From this expression, Kurul \& Podowski \cite{kurul_multidimensional_1990} assumed that the bubble growth time is negligible versus the bubble wait time. However, we question this hypothesis in the light of experimental data in the following subsection.

\subsection{Bubble wait time}

After departure from its nucleation site, cold liquid is displaced towards the wall due to departing bubble movement. From there, a transient conductive heat transfer occurs and delays the nucleation of a new bubble. This time gap is called the bubble wait time and is a crucial parameter in Heat Flux Partitioning.
%since it scales both the transient heat transfer to the liquid and the \green{bubble departure frequency}.

\begin{figure}[H]
\centering
\includegraphics[width=0.75\linewidth]{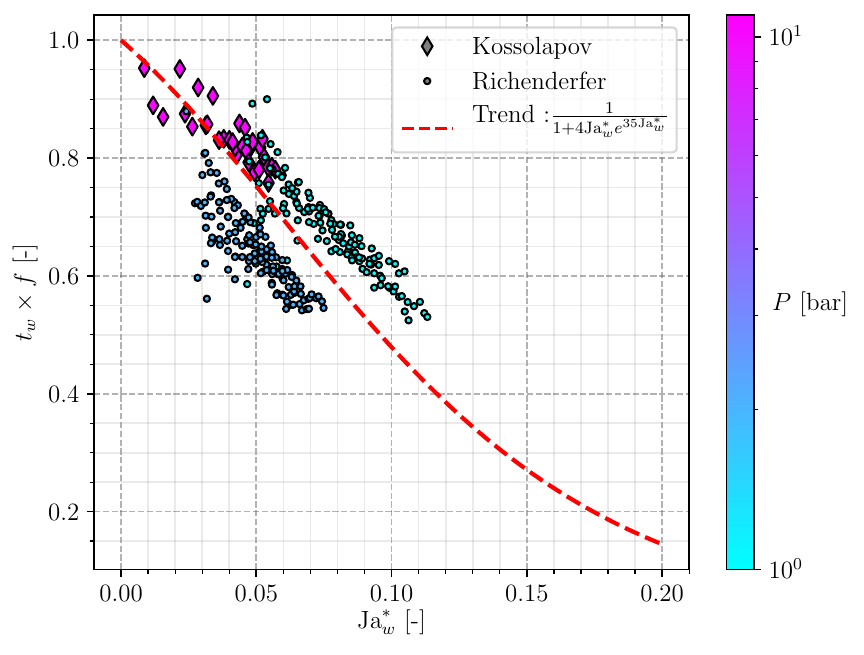}
\caption{Wait time data}
\label{fig:tw_data}
\end{figure}

First approaches considered that the bubble wait time was very large versus bubble growth time, thus neglecting the latter \cite{cole_bubble_1967, kurul_multidimensional_1990}. However, though few wait time measurements in vertical flow boiling are available in the literature (those existing to the best of our knowledge are gathered on Table \ref{tab:tw_exp_data}), we can still use the data provided by Richenderfer \cite{richenderfer_investigation_2018} and Kossolapov \cite{kossolapov_experimental_2021} to assess the validity of this hypothesis by visualizing the product $t_{w} f$ in their experiments, indicating the proportion of wait time over a full nucleation cycle (Figure \ref{fig:tw_data}).

%by Basu et al. \cite{basu_wall_2005}, 

\begin{table}[h!]

%\begin{changemargin}{-1cm}{0cm}

\noindent\makebox[\textwidth]{

\scriptsize
\centering
\begin{tabular}{p{20mm}|c c c c c c c} 
Author & Fluid & $P$ [bar] & $G_{L}$ [$\debm$] & $\Delta T_{L}$ [K] & $\phi_{w}$ [MW/m\up{2}] & $\Delta T_{w}$ [K] & $t_{w}$ [ms] ($N_{mes}$)\\
\hline
\\

Basu \etal \cite{basu_wall_2005} \newline (2005) & Water & 1.01 & 346.0 & 8.35 - 46.5 & N.A. & 9.83 - 17.5 & 0.797 - 13.3 (19) \\
\\

Richenderfer \cite{richenderfer_investigation_2018} \newline (2018) & Water & 1 - 2 & 1000 - 2000 & 5 - 20 & 0.74 - 7.13 & N.A. & 0.914 - 6.02 (259) \\
\\

Kossolapov \cite{kossolapov_experimental_2021} \newline (2021) & Water & 10.5 & 500 - 2000 & 10 & N.A. & 0.12 - 25.9 & 6.13 - 85.9 (33) \\
\\
\hline
\end{tabular}
}

\caption{Bubble wait time data in vertical flow boiling. Wall superheat values for Richenderfer data are estimated using Frost \& Dzakowic correlation \cite{frost_extension_1967}. For Basu \etal data, \green{bubble departure frequency} was not provided parallel to wait time measurements.}
\label{tab:tw_exp_data}

%\end{changemargin}

\end{table}

Figure \ref{fig:tw_data} shows that wait time amplitude can span from nearly 100\% down to 50\% of the nucleation period at large wall superheat, which demonstrates that flow conditions will largely influence the competition between wait and growth times. Therefore, strong assumptions regarding the scaling of $t_{g,d}$ over $t_{w}$ (negligibility \cite{kurul_multidimensional_1990} or constant proportionality \cite{zhou_mechanistic_2021}) will surely lack of generic physical behavior. As an example, a simple empirical trend is proposed to approximate the wait time proportion in the nucleation cycle:

\begin{equation}
    t_{w} f = \dfrac{1}{1+4\Ja_{w}^{*} e^{35 \Ja_{w}^{*}}}
\end{equation}

 This formulation consistently equals 1 at zero wall superheat (no nucleation), and tends to 0 when wall superheat tends to infinity (infinitely fast thermal boundary layer reconstruction). However, experimental measurements at higher values of $\Ja_w^*$ are clearly missing to fully validate the proposed sigmoid-shaped behavior.

\npar
As a consequence of those observations, we chose to compute the wait time using a more physical formulation to try to capture its variation with thermal-hydraulic conditions. After testing multiple expressions from the literature, we chose to use the expression of Yeoh \etal \cite{yeoh_fundamental_2008} who proposed a law that accounts for the contact angle value:

\begin{align}
t_{w} &= \frac{1}{\pi \eta_{L}}\crocht{ \frac{\parth{\Delta T_{L} + \Delta T_{w}} C_{1} R_{c}}{\Delta T_{w} - \dfrac{2\sigma T_{sat}}{C_{2}\rho_{V}h_{LV}R_{c}}}  }^{2}
\label{eq:twait_yeoh}\\
C_{1} &= \frac{1 + \cos{\theta} }{\sin{\theta}}\ ;\ C_{2} = \frac{1}{\sin{\theta}}
\end{align}

This requires an estimation of the cavity radius $R_{c}$. Yeoh \etal \cite{yeoh_fundamental_2008} also proposed an expression for $R_c$ that also depended on the contact angle $\theta$, but it was found out that the resulting model was way too sensitive to its value. That is why we chose to use the expression of Han \& Griffith which is independent of $\theta$ \cite{chi-yeh_mechanism_1965}:

\begin{align}
R_{c} &= \frac{2\sigma T_{sat}}{\rho_{V}h_{LV}\Delta T_{w}}
\label{eq:rc_han}\\
\end{align}

The selected expressions are confronted to the empirical formulation of Basu \etal \cite{basu_wall_2005}:

\begin{equation}
t_{w} = 139.1\Delta {T_{w}}^{-4.1}
\end{equation}

\begin{figure}[H]
\centering
\includegraphics[width=0.49\linewidth]{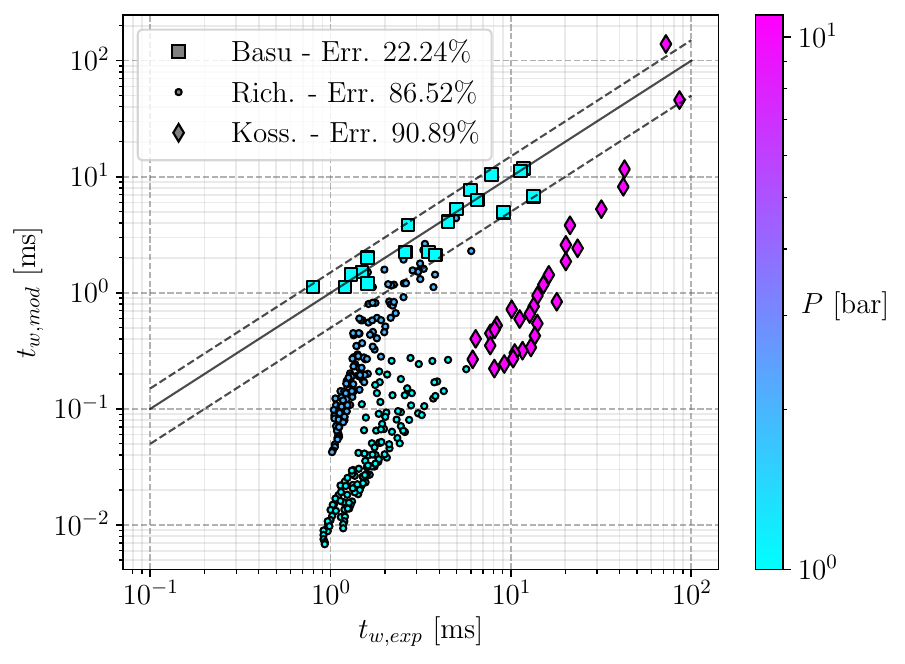}
\includegraphics[width=0.49\linewidth]{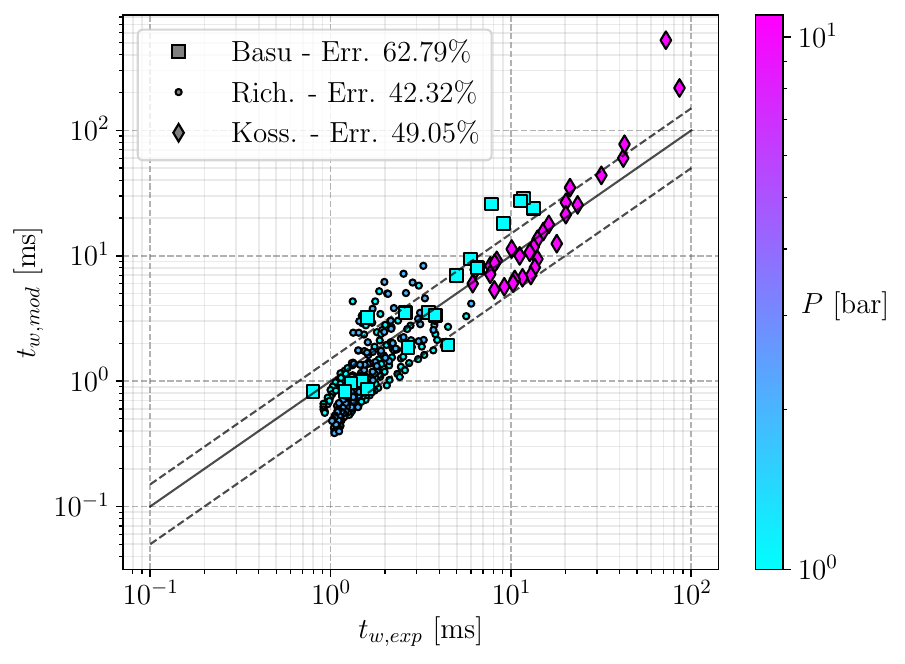}
\caption{Wait time predictions for Basu \etal (left) and selected formulation (right), $\pm 50\%$ error bars in dashed lines.}
\label{fig:tw_comp}
\end{figure}

Figure \ref{fig:tw_comp} shows that the chosen formulation is able to predict the experimental data from Table \ref{tab:tw_exp_data} with an average error around 50\%, while Basu \etal formulation fails to reproduce the wait times from Richenderfer and Kossolapov due to its empirical nature correlated on their own wait time data.

Here again, we have to acknowledge that an average error of $50\%$ on the wait time shall have a large influence on the final HFP results. However, it currently seems hard to ensure better reproduction of the available experimental results using other existing formulations of the wait time. 

\subsection{Bubble growth time}

As discussed in Favre \etal \cite{favre_updated_2023}, the bubble growth both before and after sliding can be acceptably reproduced using a heat diffusion law similar to those of Plesset \& Zwick \cite{plesset_growth_1954} or Scriven \cite{scriven_dynamics_1959}. Therefore, since the bubble growth is modeled as:

\begin{equation}
    R\parth{t} = K \Ja_w \sqrt{\eta_L t}
\end{equation}

Then the bubble growth time reads:

\begin{equation}
    t_{g} = \parth{\dfrac{R}{K \Ja_w}}^2
\end{equation}
where it corresponds to the time until departure if $R=R_d$, or lift-off by sliding coalescence if $R=R_{sl}$.

The value of the growth constant $K$ remains an empirical constant of the model that will have to be chosen consistently with the considered flow boiling conditions (\eg it should theoretically not reach values much larger than 2 \cite{favre_updated_2023}). 

\subsection{Quenching time}

The duration over which quenching heat transfer occurs can vary depending on the flow conditions. In particular, if the wait time between two bubble nucleation events is long, heat transfer during this period can switch from transient heat transfer contributing to boundary layer reconstruction back to forced convection. As introduced by Basu \etal \cite{basu_wall_2005}, equating the heat transfer coefficients in Eq. \ref{eq:phi_cL} and \ref{eq:phi_q} allows to express the maximum quenching time $t^*$ after which forced convection will overcome transient heat transfer due to quenching:

\begin{equation}
    t^* = \parth{\dfrac{\lambda_L}{h_{c,L}}}^2 \dfrac{1}{\pi \eta_L}
\end{equation}

Then, depending on the relationship between $t^*$ and $t_w$, we can define the quenching time as:

\begin{equation}
    t_q = \mathrm{min}\parth{t^*,t_w}
\end{equation}

\section{Nucleation site density and bubble interactions}
\label{sec:bubble_interactions}

\subsection{Nucleation site density correlation}
\label{subsec:NSD}

One of the firstly identified behavior of the NSD was its power dependency with the wall superheat ($N_{sit} \propto {\Delta T_{w}}^{m}$), which is adopted in the well-known correlation of Lemmert \& Chawla \cite{lemmert_influence_1977} : 

\begin{align}
N_{sit}=\crocht{210\parth{T_{w}-T_{sat}}}^{1.8}
\label{eq:nsit_lemmert}
\end{align}

However, the nucleation site density is intrinsically linked to the heater material properties and is also proven to be largely influenced by the operating pressure \cite{borishanskii_heat_1969, hibiki_active_2003, kommajosyula_development_2020}. Therefore, Hibiki \& Ishii \cite{hibiki_active_2003} later proposed an upgraded formulation of the nucleation site density that includes the effect of pressure. Although this correlation largely improved the prediction of experimental results, their mathematical expression still presented issues with diverging values when reaching large wall superheat, making it unsuitable for large $\Ja_w$ cases. 

To mitigate this issue, a more recent expression of $N_{sit}$ has been proposed by Li \etal in 2018 \cite{li_development_2018}. It was validated over a large range of measurements by including a more realistic power law for $\Delta T_{w} = T_w - T_{sat}$ avoiding the divergence at high pressure and superheat. They also added the effect of contact angle and its evolution with temperature \eg its decrease close to 0 $\degree$ when approaching the critical temperature \cite{song_temperature_2021}:

\begin{align}
N_{sit} &= N_{0}e^{\mathrm{f}\parth{P}} {\Delta T_{w}}^{A\Delta T_{w} + B} \parth{1- \cos{\theta}}
\label{eq:nsit_li}\\
\mathrm{f}\parth{P} &= 26.006 - 3.678 e^{-2P} - 21.907e^{-P / 24.065}\\
A &= -2\times 10^{-4} P^{2} + 0.0108P + 0.0119\\
B &= 0.122P +1.988\\
1-\cos{\theta} &= \parth{1-\cos{\theta_{0}}}\parth{\frac{T_{c}-T_{sat}}{T_{c}-T_{0}}}^{\gamma}
\end{align}
with $P$ in MPa, $\theta_{0}$ the contact angle at room temperature $T_{0}$, and default value being for water $N_0 = 1000$m\up{-2},$\theta_{0}=41.37 \degree$, $T_{c}=374 \degree$C $T_{0}=25\degree$C, $\gamma = 0.719$.

In reality, the value of $\theta_0$ and $T_0$ in this expression should be adapted to realistic values depending on the considered fluid / heater material characteristics.

Later, Zhou \etal \cite{zhou_experimental_2020} also proposed their own NSD correlation based on their own measurements and those of Basu \etal \cite{basu_onset_2002} by including the effect of pressure and contact angle.

In order to compare the predictive capability of these models, we gathered NSD experimental measurements from the literature that are summed up in Table \ref{tab:nsit_exp_data}. Experiments cover pressures from 1.01 bar to 198 bar for boiling water. The value of $\theta_0$ is adapted to realistic values regarding the experiments (\eg large contact angle for Richenderfer \cite{richenderfer_investigation_2018} and Kossolapov \cite{kossolapov_experimental_2021} experiments where they use very smooth non-prototypical surface with ITO deposit).

\begin{table}[!ht]

%\begin{changemargin}{-1cm}{0cm}

\noindent\makebox[\textwidth]{

\scriptsize
\centering
\begin{tabular}{p{20mm}|c c c c c c c c} 
Author & Fluid &  $P$ [bar] & $G_{L}$ [$\debm$] & $\Delta T_{L}$ [K]  & $\Delta T_{w}$ [K] & $\theta_{0}$ [$\degree$] &  $N_{mes}$ [-]\\
\hline
\\

Zhou \cite{zhou_experimental_2020} \newline (2020) & Water & 1.21 - 3.12 & 482.7 - 1930.6 & 8 - 15  & 6.7 - 20.2 & $51$ & 60 \\
\\

Richenderfer \cite{richenderfer_investigation_2018} \newline (2018) & Water & 1.01 & 500 - 1000 & 10 & 21.7 - 42.8 & $80$ & 49 \\
\\

Kossolapov \cite{kossolapov_experimental_2021} \newline (2021) & Water & 1.01 - 75.8 & 500 - 2000 & $80$ &10 & 80 & 63 & \\
\\

Borishanskii \cite{borishanskii_heat_1969} \newline (1966) & Water & 1.01 - 198 & N.A. & N.A. & 1.75 - 17.3 & $45$ & 132 \\
\\

\hline
\end{tabular}
}

\caption{Selected NSD data in flow boiling}
\label{tab:nsit_exp_data}

\end{table}

\begin{figure}[H]
\centering
\subfloat[Lemmert \& Chawla model]{
\includegraphics[width=0.49\linewidth]{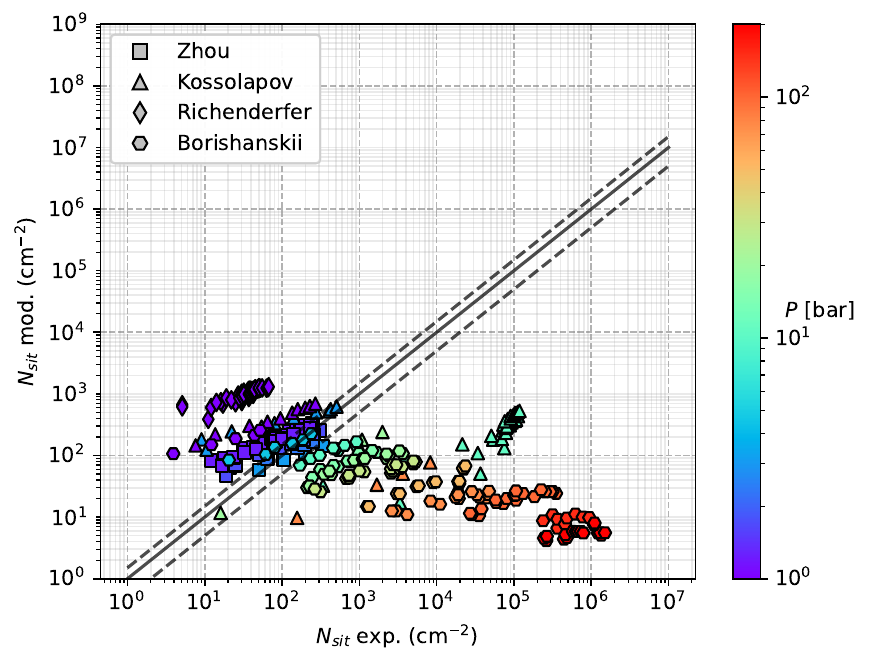}
} 
\subfloat[Zhou model]{
\includegraphics[width=0.49\linewidth]{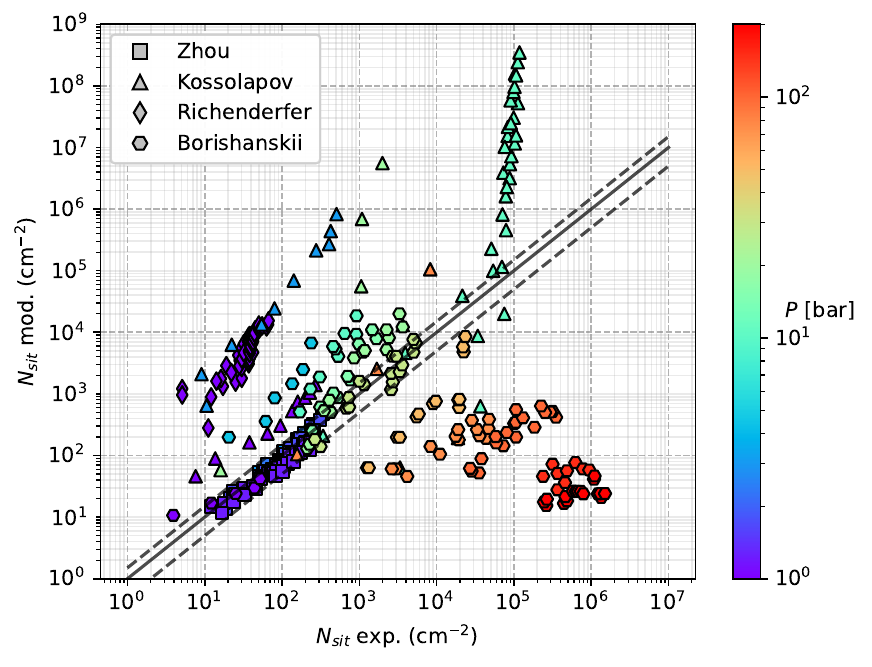}
}

\subfloat[Hibiki \& Ishii model]{
\includegraphics[width=0.49\linewidth]{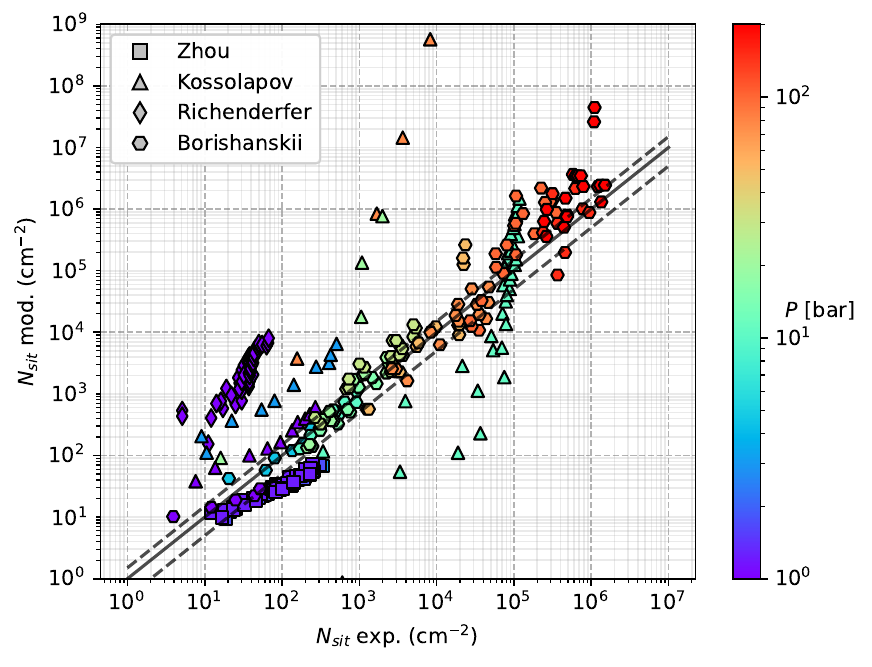}
}
\subfloat[Li \etal model]{
\includegraphics[width=0.49\linewidth]{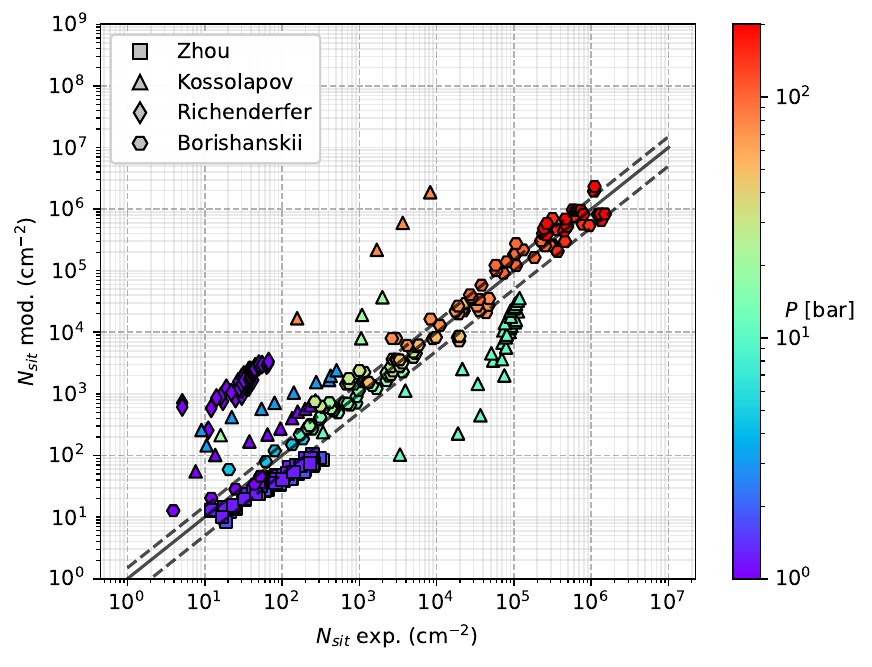}
}

\caption{Nucleation site density predictions using various correlations with $\pm 50\%$ error bars indicated in dashed lines.}
	
\label{fig:nsd_comp}	
\end{figure}

Figure \ref{fig:nsd_comp} presents the results obtained of the NSD database for each of the aforementioned models.

As already discussed, Lemmert \& Chawla formulation clearly fails to capture the pressure dependency of the experimental data. Zhou \etal formulation naturally performs well on their own data, managing to perform correctly for lower pressure experiments but showing larger discrepancies otherwise. Finally, Hibiki \& Ishii and Li \etal models both perform quite well by significantly reproducing the influence of pressure over the whole dataset, although we see that Li \etal model definitely avoids very large overestimation when reaching the largest wall superheat for Kossolapov and Borishanskii data. Overall, it seems that all models are struggling to precisely predict Richenderfer and Kossolapov data, which can be explained by the fact that they use prototypical surfaces, \blue{in} which nucleation site density is much lower than industrial metal ones. \blue{An interesting perspective here would be a systematic investigation of the influence of the wall roughness and topology over the nucleation site density in order to be able to include it in future models \eg rougher surfaces being expected to present much more nucleation sites than very smooth ones \cite{gorenflo_bubble_2004, kotthoff_heat_2006}. One could also try to connect wall roughness to cavity size and radius, playing a direct but very complicated role in the boiling process \cite{hsu_size_1962}}.

In the end, this comparison shows that achieving a proper prediction of the nucleation site density is very complicated if one wants to cover a broad range of flow conditions and heater materials. Acknowledging this difficulty, Li \etal model still seems to be the best performing one and will thus be selected for the modeling.

\subsection{Nucleation site static suppression}

Although we have access to an evaluation to the number of available sites for nucleation, it still has to be determined wether all of those sites will actually contribute to bubble generation \ie determining the active nucleation site density $N_{sit,a}$. Indeed, if we consider the product $\pi R_{d}^2 N_{sit}$, its value must remain lower than one at all time. However, written this way, nothing prevents it to reach nonphysical values greater than 1 meaning that the sum of bubbles projected areas represent more than 100\% of the wall surface. That is why Kurul \& Podowski \cite{kurul_multidimensional_1990} chose to clip its value to 1 if it ever was calculated to be larger. 

A finer approach to this question has been proposed by Gilman \& Baglietto \cite{gilman_self-consistent_2017}. In their work, they estimate the number of sites that would be overlapped by a neighboring growing bubble, thus impeding the site to nucleate a bubble itself and then being deactivated.

\begin{figure}[H]
\centering
\includegraphics[width=0.9\linewidth]{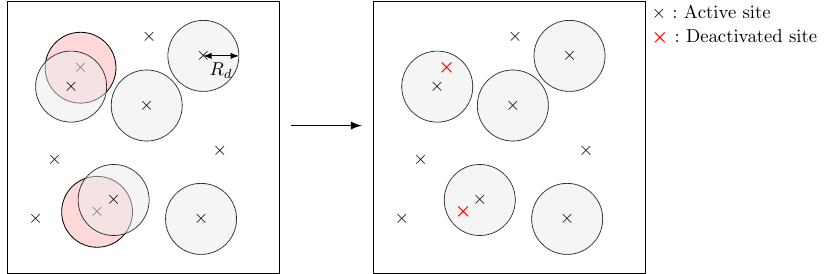}
\caption{Sketch of the geometrical overlapping leading to static deactivation. Bubbles in red can not be accommodated on the surface due to their site laying below an existing bubble.}
\label{fig:static_deactivation}
\end{figure}

 We will follow the same approach here: let us consider the nucleation site density $N_{sit}$ computed by a correlation as in Subsection \ref{subsec:NSD}. As mentioned earlier, if we distribute a given number of bubbles of radius $R_{d}$ over the different sites, we have no guarantee that the correlation avoids too large values of $N_{sit}$ that would lead to geometrical overlapping as shown on Figure \ref{fig:static_deactivation}.

Thus, we need a correction of $N_{sit}$ to obtain the actual number of active sites $N_{sit,a}$ that can geometrically fit on the surface regarding the nucleation parameters. Given a bubble growth time before departure $t_{g,d}$ and an average \green{bubble departure frequency} $f$, we can estimate the actual number of bubbles growing attached to their sites on the boiling surface as:

\begin{equation}
N_{b} = t_{g,d}\times f \times N_{sit,a}
\label{eq:bubble_density}
\end{equation}

 Experimental observations have showed that the active nucleation site density was likely to follow a space distribution close to a Poisson point process \cite{gaertner_population_1960, sultan_spatial_1978, del_valle_subcooled_1985, zhou_experimental_2020-1}. Therefore, assuming this stochastic behavior, we can use $N_{b}$ as an event density in a spatial Poisson point process probability density function to estimate the probability to have an undesired overlapping \ie two simultaneous bubbles of radius $R_{d}$ on neighboring sites at a distance $r\leq R_{d}$:

\begin{align}
\mathcal{P}\parth{r\leq R_{d}} &=1-\underbrace{ \mathcal{P}\parth{N\parth{\pi R_{d}^{2}}=0} }_{\text{No bubble nucleates within $\pi R_{d}^{2}$}} = 1-e^{-N_{sit,a}t_{g,d}f\pi R_{d}^{2}} = \mathcal{P}
\end{align}

This overlapping probability $\mathcal{P}$ can then be used to mitigate the number of sites given by the NSD correlation, yielding:

\begin{align}
&N_{sit,a}=\parth{1-\mathcal{P}}N_{sit} \\
\Leftrightarrow\  &N_{sit,a}t_{g,d}f\pi R_{d}^{2}e^{N_{b}t_{g,d}f \pi R_{d}^{2}}= N_{sit}t_{g,d}f \pi R_{d}^{2} \\
\Leftrightarrow\   &N_{sit,a} = \frac{\mathcal{W}\parth{N_{sit}{A_{sit}}}}{{A_{sit}}}
\label{eq:static_deact}
\end{align}
where $A_{sit}=t_{g,d}f \pi R_{d}^{2}$ and $\mathcal{W}$ is Lambert's W-function (reciprocal of $x \rightarrow xe^{x}$).

The evaluation of $\mathcal{W}$ can easily be achieved with a few iterations of a bisection method. Otherwise, Kommajosyula proposed an approximation to allow its direct computation \cite{kommajosyula_development_2020}.

\npar

\begin{figure}[H]
\centering
\subfloat[Li \etal correlation (Eq. \ref{eq:nsit_li})]{
\includegraphics[width=0.5\linewidth]{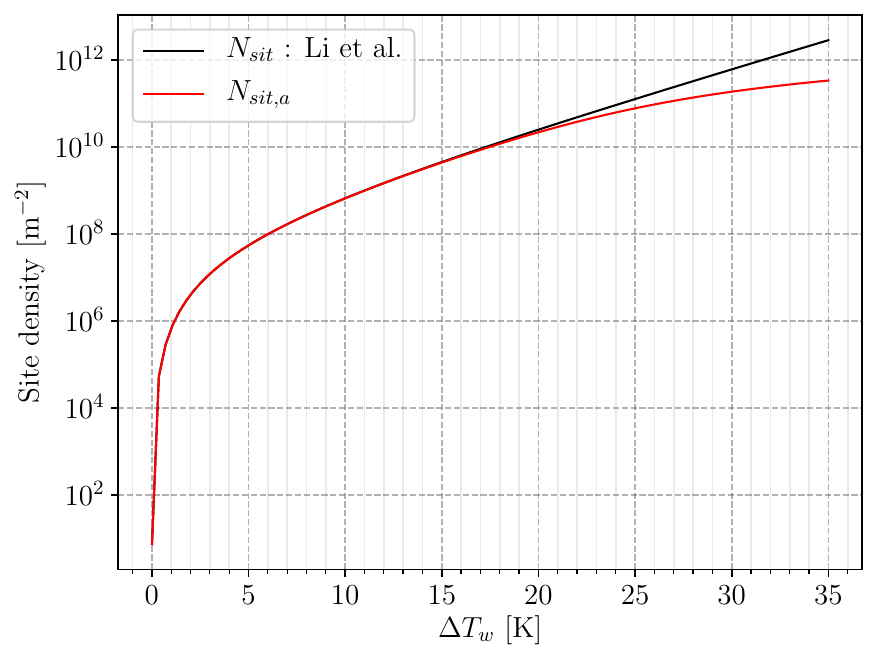}
}
\caption{Static deactivation correction tested with Li \etal NSD correlation for water at 40 bar, $f=200$~Hz, $t_{g,d} = 0.1$~ms, $\theta=80\degree$ and $R_{d} = 0.01$~mm.}
\label{fig:nsit_CSR}
\end{figure}

On Figure \ref{fig:nsit_CSR} we show the impact of the correction of Eq. \ref{eq:static_deact} on the NSD correlation of Li \etal (Eq. \ref{eq:nsit_li}).

Initially, this correction proposed by Gilman \& Baglietto was meant to mitigate the diverging behavior of Hibiki \& Ishii NSD law. However, in our case, Figure \ref{fig:nsit_CSR} shows that even Li \etal correlation will be mitigated when reaching large values of wall superheat depending on the boiling parameters $f$, $R_d$ and $t_{g,d}$, which then ensures that the number of active nucleation sites will remain physically consistent for the boiling surface.

\subsection{Static bubbles coalescence}

\begin{figure}[H]
\centering
\includegraphics[width=0.5\linewidth]{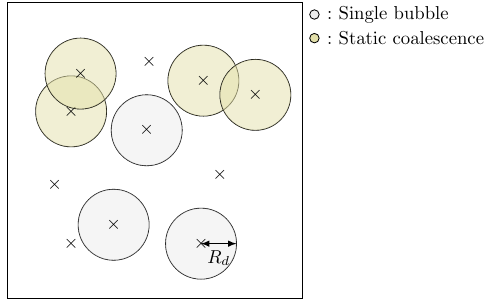}
\caption{Sketch of the static coalescence phenomenon.}
\label{fig:static_coal}
\end{figure}

After identifying the actual number of bubble-generating sites have been identified, we want to consider other interaction phenomena that can occur on the boiling surface. For instance, if two bubbles are simultaneously growing on sites at a distance lower than $2R_{d}$, the bubbles will coalesce while growing up to the detachment diameter (Figure \ref{fig:static_coal}). 

Following a reasoning analogous to the derivation of the active NSD $N_{sit,a}$ in previous subsection, we use $N_b$ as the event density of a spatial Poisson point process to compute the probability of static coalescence:

\begin{align}
\mathcal{P}\parth{r \leq 2R_{d}} &=\int_{0}^{2R_{d}}f\parth{r} \mathrm{d}r = 1 - e^{-N_{b} \pi \parth{2R_{d}}^{2}}=\mathcal{P}_{coal,st}
\label{eq:proba_coal_st}
\end{align}

The density of bubble-generating sites that will lead to a static coalescence can then be estimated as :

\begin{align}
N_{coal,st}=\mathcal{P}_{coal,st}N_{sit,a}
\label{eq:sites_static_coal}
\end{align}

\green{This value can be interpreted in two equivalent ways, either as single bubble nucleation or pairs of bubbles coalescing:
\begin{itemize}
    \item There are $N_{coal,st}$ sites that will generate bubbles up to a radius $R_d$ before coalescing with another one of the same size nearby.
    \item There are $N_{coal,st}/2$ sites that will generate a coalesced bubble of radius $R_{coal,st}=\parth{R_d^3 + R_d^3}^{1/3}$, \ie two sites are needed to generate one coalesced bubble out of two single ones.
\end{itemize}}

\begin{figure}[H]
\centering
\includegraphics[width=0.6\linewidth]{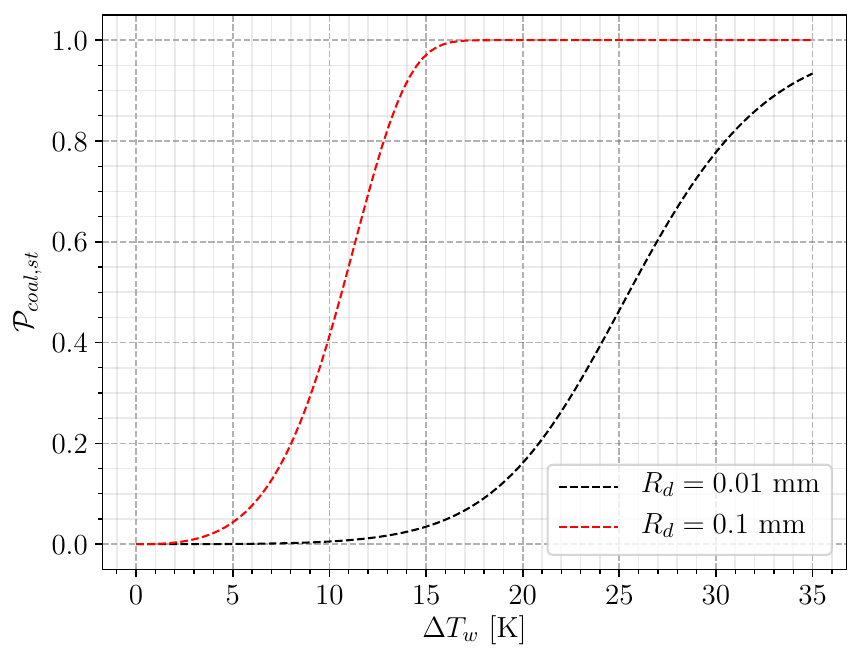}
\caption{Static coalescence probability with Li \etal NSD correlation for water at 40 bar, $f=200$~Hz, $t_{g,d} = 0.1$~ms and $\theta=80\degree$.}
\label{fig:pcoal_st}
\end{figure}

Figure \ref{fig:pcoal_st} presents the evolution of $\mathcal{P}_{coal,st}$ with the wall superheat for two departure radius values, using the same conditions as in Figure \ref{fig:nsit_CSR}. The probability of static coalescence consistently increase with the bubble departure radius and with the wall superheat, reaching regimes where $\mathcal{P}_{coal,st} = 1$ meaning the boiling surface will be saturated with bubbles coalescing while nucleating close to each other at large wall temperature.

This finally allows to compute the static coalescence \green{evaporation heat flux} that was considered as one of the heat transfer mechanisms to account for in the HFP model (Eq. \ref{eq:phie_st}). Still, we must acknowledge that this calculation should slightly overestimate the number of static coalescence sites since we consider these events to occur at bubble departure radius. However, since bubbles are likely to quickly depart by sliding in vertical flow boiling, we expect this overestimation to have a limited impact over the final results.

\subsection{Sliding bubbles coalescence}
\label{subsec:sliding_coal}

Now that we have access to the number of sites leading to static coalescence, we can deduce that the remaining sites will be generating sliding bubbles. Since we considered that sliding bubbles will lift-off by coalescing with a static bubble on its site with a radius close to $R_d$, we have half the remaining number of sites that will nucleate sliding bubbles $N_{sl}$ and the other half will nucleate bubbles that will stay on their site before coalescing with a sliding bubble. The total number of sliding coalescence nucleation sites $N_{coal,sl}$ (Eq. \ref{eq:phie_sl}) is then:

\begin{equation}
    N_{sl} = N_{coal,sl} = \dfrac{1}{2}\parth{N_{sit,a} - N_{coal,st}} = \dfrac{1}{2}\parth{1-\mathcal{P}_{coal,st}}N_{sit,a}
\end{equation}

Therefore, the sliding coalescence hypothesis implies that the distance covered by sliding bubbles should be the average distance between two bubbles nucleating on their site $s_b$. This quantity is easily calculated by relying on the spatial Poisson point process giving the average distance between two events:

\begin{equation}
    l_{sl} = s_b = \dfrac{1}{2\sqrt{N_b}}
\end{equation}

Other HFP models sometimes rely on the average distance between two active nucleation sites to estimate sliding distances \cite{basu_wall_2005, kommajosyula_development_2020}, but this choice would result in quasi null $l_{sl}$ values at high superheat or high pressure when the nucleation site density becomes very large which is questionable since experiments have observed bubble sliding over significant distances even under high pressure conditions \cite{kossolapov_experimental_2021}.

One should note that we did not account for nucleation site deactivation due to sliding bubbles nor for coalescence between two sliding bubbles. Doing so would require to be able to compute a reference sliding length $l_{sl,0}$ after which single bubbles will lift-off by themselves, which is still an open question in the case of vertical flow boiling as discussed in Section \ref{sec:hfp_model}.

\section{Heat transfer areas and remaining parameters}
\label{sec:HT_areas}

\subsection{Single bubble sliding area}

The estimation of the bubble sliding area is crucial since it directly scales the heat flux partitioning through the wall area uninfluenced by bubble presence. The question of whether the impacted area is larger than the projected area of the bubble is still an open question, but recent results from Kossolapov \cite{kossolapov_experimental_2021} indicated that the bubble thermal footprint seems to remain within the bubble projected area, which will then be our assumption here. Usually, most HFP models compute the bubble sliding area as \cite{kommajosyula_development_2020, gilman_self-consistent_2017, basu_wall_2005}:

\begin{equation}
A_{q,1b} = l_{sl}\parth{R_{d} + R_{lo}}
\end{equation}

\begin{figure}[H]
\centering
\subfloat[$l_{sl}< R_{lo}-R_{d}$]{
\includegraphics[width=0.15\linewidth]{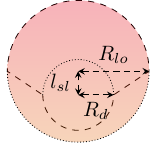}
}
\hfill
\subfloat[$R_{lo}-R_{d} \leq l_{sl}$ and $l_{sl} \leq R_{lo} + R_{d}$]{
\includegraphics[width=0.15\linewidth]{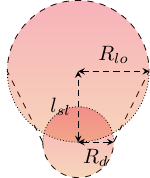}
}
\hfill
\subfloat[$l_{sl} > R_{lo} + R_{d}$]{
\includegraphics[width=0.15\linewidth]{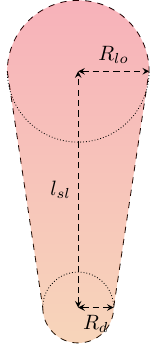}
}
\caption{Quenching area shape depending on the relation between $R_{d}$, $R_{lo}$ and $l_{sl}$.}
\label{fig:slide_area}
\end{figure}

However, depending on the magnitudes of $l_{sl}$, $R_{d}$ and $R_{lo}$, the quenching area for a single sliding bubble may vary in shape as shown on Figure \ref{fig:slide_area}.

\npar

Based on Figure \ref{fig:slide_area}, we have:

\begin{align}
A_{q,1b} = &
\begin{dcases}
\pi R_{lo}^{2} & \text{if } l_{sl}\leq R_{lo}-R_{d} \\
\frac{1}{2}\pi R_{d}^{2} + l_{sl}\parth{R_{d}+R_{lo}} + \frac{1}{2}\pi R_{lo}^{2} & \text{if } l_{sl} \geq R_{lo}+R_{d}
\end{dcases}
\label{eq:Aq_1b}
\end{align}

For the intermediate case when $R_{lo}-R_{d}\leq {l_{sl}} \leq R_{lo} + {R_{d}}$, we simply linearly interpolate those two expressions. In the case of sliding coalescence, the lift-off diameter $R_{lo}$ will be the final diameter of the merged bubbles $\parth{R_d^3 + R_{sl}^3}^{1/3}$.

The sliding diameter before coalescence $R_{sl}$ is computed by iteratively solving the bubble sliding until it reaches the sliding distance $l_{sl} = s_b$ as discussed in Subsection \ref{subsec:sliding_coal} 
\subsection{Quenching area}

Knowing the wall area swept by sliding bubbles, and assuming that quenching due to static bubble coalescence will occur over the two initial bubbles' projected area, we can write:

\begin{equation}
    A_q = N_{coal,st} \pi R_d^2 + N_{sl} A_{q,1b}
\end{equation}

\subsection{Vapor convective area}

Using the bubble growth law, we can compute the average size of a bubble over its lifetime:

\begin{equation}
    \left<R\right> = \dfrac{1}{t_g} \int_{0}^{R} r\ \mathrm{d}r =  \dfrac{1}{t_g} \int_{0}^{t_g} K \Ja_w \sqrt{\eta_l t }\ \mathrm{d}t= \dfrac{2}{3} R
\end{equation}

The wall area in direct contact with vapor through bubbles' base can be expressed (assuming bubbles with a truncated sphere shape respecting the contact angle $\theta$):

\begin{equation}
    A_{c,V} = \parth{N_{coal,st} + \dfrac{N_{coal,sl}}{2}} \pi  \parth{\dfrac{2}{3}R_d \sin{\theta}}^2 \times t_{g,d}f + \dfrac{N_{coal,sl}}{2} \pi \parth{\dfrac{2}{3} R_{sl} \sin{\theta}}^2 \times t_{g,lo}f
\end{equation}

\subsection{Liquid convection area}

From the previous expressions, we can finally deduce the area undergoing liquid convection:

\begin{equation}
    A_{c,L} = 1 - A_q t_q f - A_{c,V}
\end{equation}

\subsection{Remaining empiricism}

Following the proposed construction of the HFP model, we are left with a set of 3 parameters which values are to be set, namely:

\begin{itemize}
    \item The contact angle $\theta$, involved in several closure laws (bubble force balance, wait time, nucleation site density, dry area), thus likely to be a sensitive parameter that should nonetheless be chosen accordingly with the fluid / heater material to be simulated. For instance, typical values for water can reach up to 85\degree\ for non-prototypical surfaces \cite{kossolapov_experimental_2021}, or lower values laying around 20\degree for metallic surfaces under high temperature conditions \cite{song_temperature_2021}. For refrigerants such as R134a, their large wettability on metallic surfaces usually tend to exhibit small contact angles down to lower than 10\degree\ \cite{zou_heating_2013}.
    \item The contact angle half-hysteresis $\mathrm{d}\theta$, coming from the force balance model (see Figure \ref{fig:bub_bdf}) used to compute $R_d$ and bubble sliding velocity \cite{favre_updated_2023}, which will mostly impact the bubble detachment radius, thus modifying the static coalescence and \green{evaporation heat fluxes}. Its value should be also chosen appropriately \ie smaller as bubble size diminishes (surface tension force increasing relative to other forces) and possibly increasing with the liquid mass flux (higher drag inducing a larger effort likely to tilt the bubble). Typical values can reach more than 10\degree\ to 20\degree\ for large bubbles (\eg atmospheric pressure flow) down to 1\degree\ and lower for very small bubbles (high pressure flow). The surface material will also greatly influence $\dtheta$ value, \eg a smooth surface is less likely to keep the contact line sticking and will let it move with bubble sliding earlier, thus inducing smaller bubble tilt compared to rough surfaces.
    \item The bubble growth rate $K$, which value should normally lay roughly in the range between 0.1 and 2 depending on the operating conditions. For instance, this parameter shall be expected to decrease as liquid subcooling or liquid mass flux increases (bubbles that are more likely to face cold liquid will have a slower growth). For example, values close to $K=1$ have been shown to fairly reproduce bubble growth profiles for boiling water at 1 and 10 bar \cite{favre_updated_2023} on Maity \cite{maity_effect_2000} and Kossolapov \cite{kossolapov_experimental_2021} data, while values between 1.5 and 2 were better fitting for 40 bar boiling cases from Kossolapov. 
\end{itemize}

\section{Model Validation and interpretation of the results}
\label{sec:validation}

\subsection{Detailed validation on Kossolapov cases}

In order to evaluate the developed HFP model, we compare its results with experimental measurements from Kossolapov \cite{kossolapov_experimental_2021}. The experiment consists of a small heating cell in an upward pressurized flow of water, where several quantities of interest were measured such as the nucleation site density, \green{bubble departure frequency}, wait time, growth time and quenching heat transfer time along with the evolution of the wall temperature against the applied heat flux. Such data are extremely valuable to validate HFP models like the one developed in this work since it allows to achieve parallel assessment of the model's wall heat flux predictions along with some of the previously discussed closure laws.

\begin{figure}[H]
\centering
\subfloat[NSD (upper left) - BDF (upper right) - Time scales (bottom left) - Length scales (bottom right)]{
\includegraphics[width=0.8\linewidth]{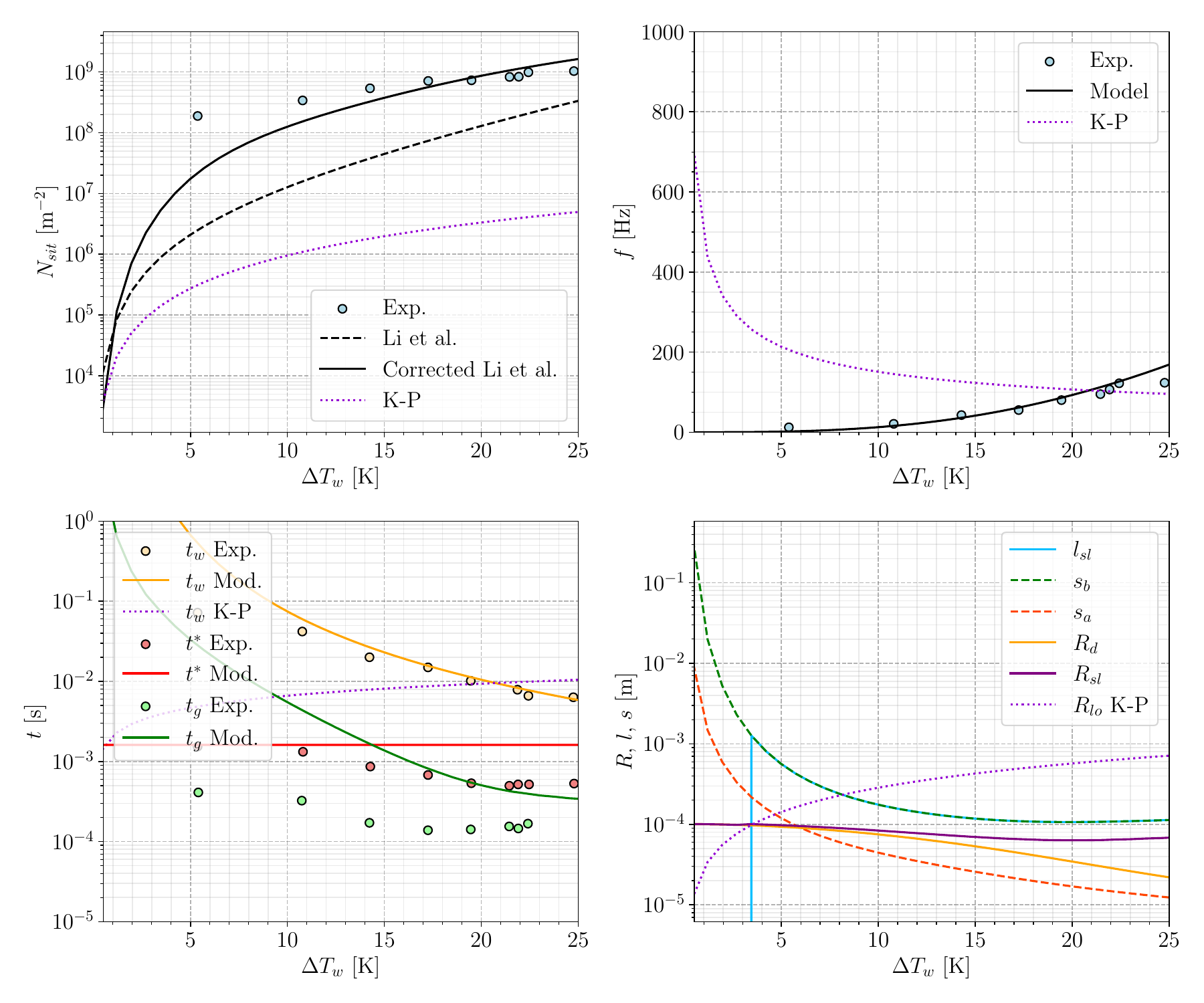}
}

\subfloat[Boiling curve - Kurul \& Podowski]{
\includegraphics[width=0.49\linewidth]{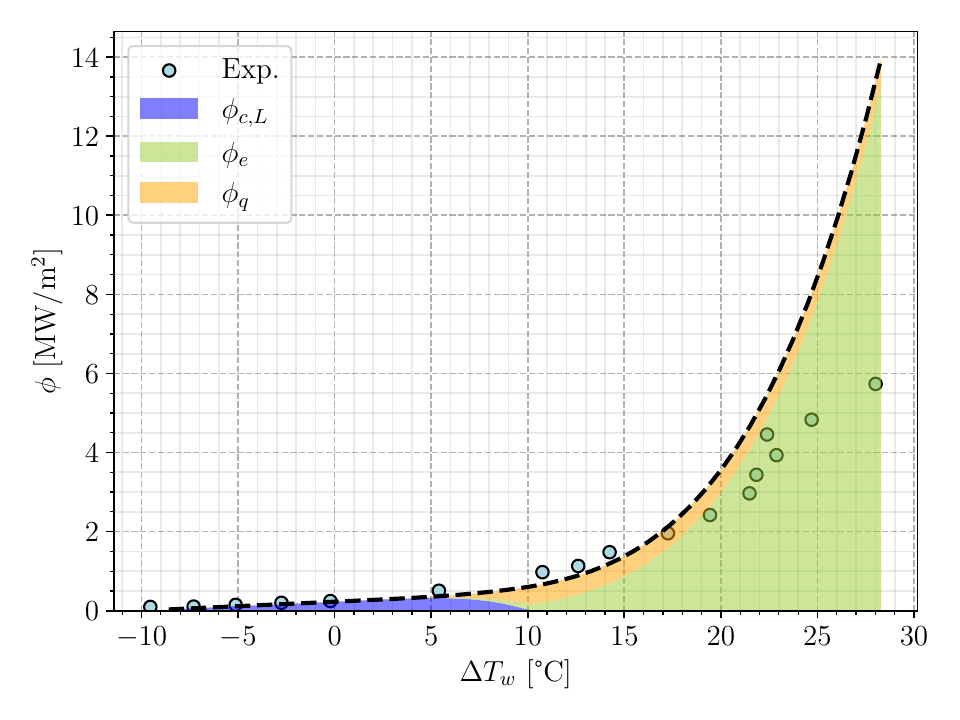}
}
\subfloat[Boiling curve - Present model]{
\includegraphics[width=0.49\linewidth]{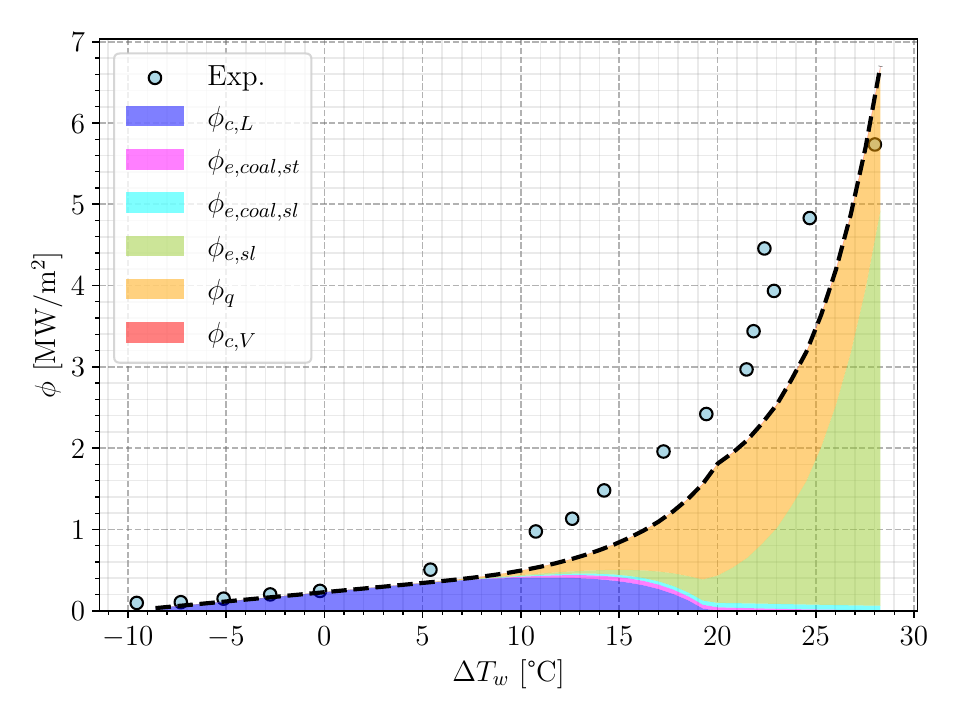}
}
\\

\caption{Kossolapov case : $P=10.5$ bar, $G=500\ \debm$, $\theta = 85 \degree$, $\dtheta = 2 \degree$, $K = 0.7$}
\label{fig:full_koss_G500}
\end{figure}

\begin{figure}[H]
\centering
\subfloat[NSD (upper left) - BDF (upper right) - Time scales (bottom left) - Length scales (bottom right)]{
\includegraphics[width=0.8\linewidth]{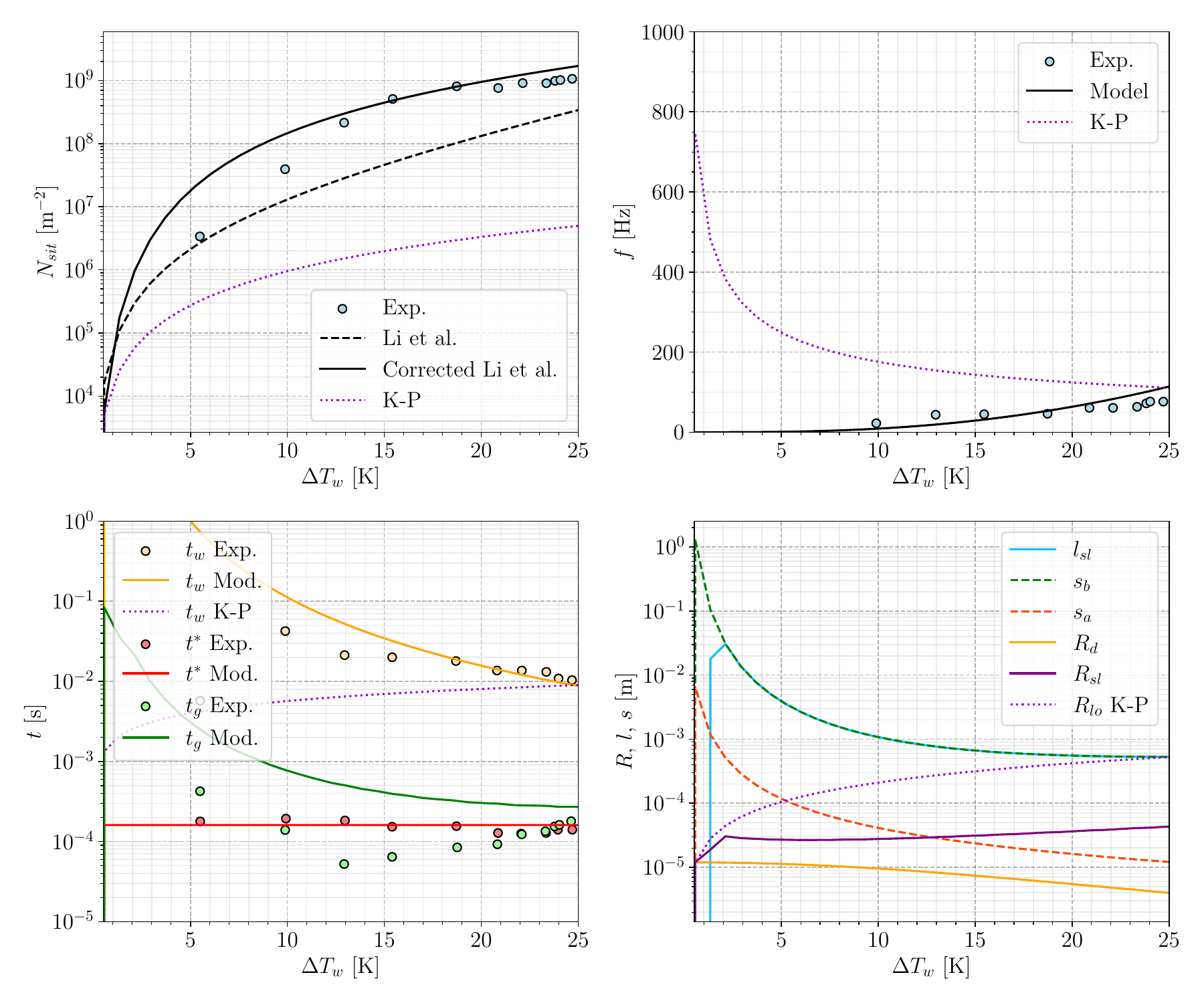}
}

\subfloat[Boiling curve - Kurul \& Podowski]{
\includegraphics[width=0.49\linewidth]{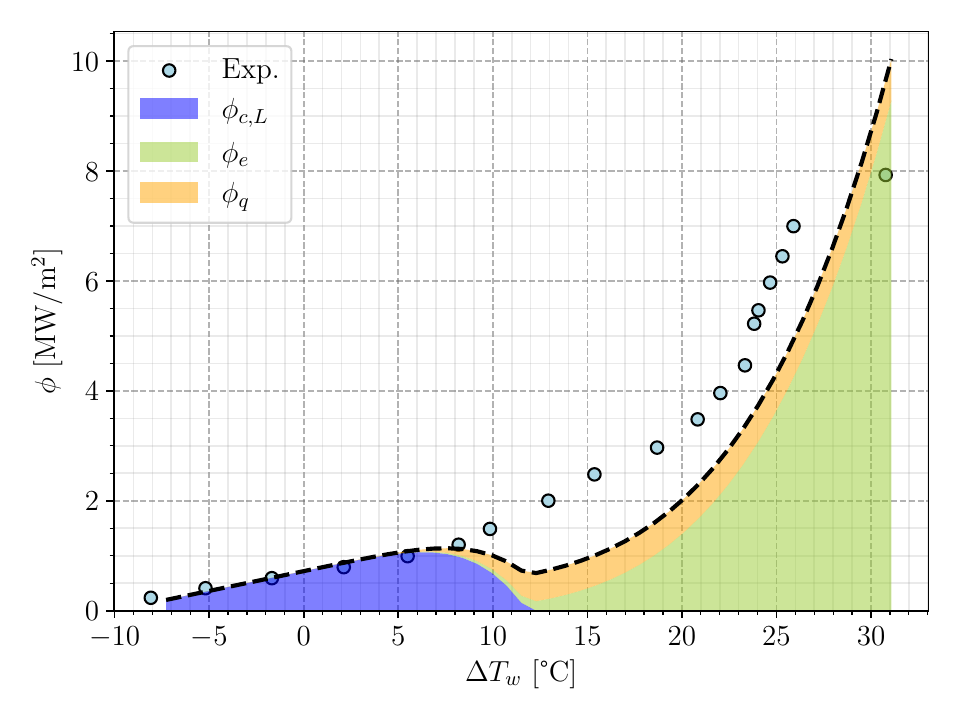}
}
\subfloat[Boiling curve - Present model]{
\includegraphics[width=0.49\linewidth]{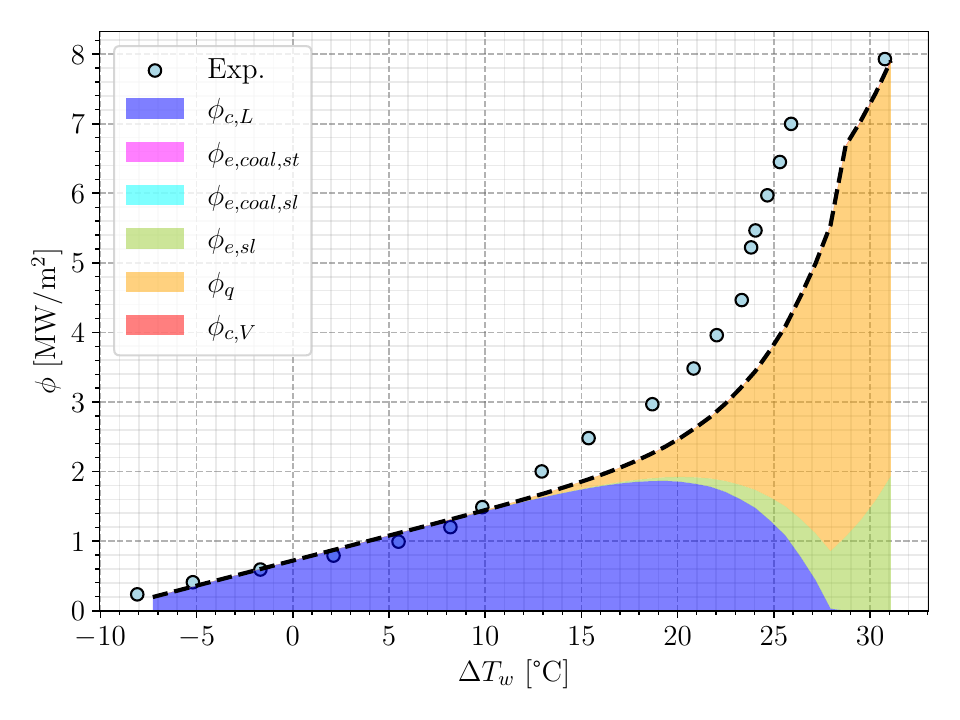}
}
\\

\caption{Kossolapov case : $P=10.5$ bar, $G=2000\ \debm$, $\theta = 85.5 \degree$, $\dtheta = 3 \degree$, $K = 0.5$}
\label{fig:full_koss_G2000}
\end{figure}

Measurements provided by Kossolapov's work are conducted at a pressure of 10.5 bar, a 10~K liquid subcooling, from which we selected two cases with liquid mass fluxes of 500 and 2000 kg/m\up{2}/s. On Figure \ref{fig:full_koss_G500} and \ref{fig:full_koss_G2000}, we present the results obtained with the proposed HFP model, including a correction applied on Li \etal \cite{li_development_2018} nucleation density law. Though the Li \etal correlation fairly captures the trend and values of the measured nucleation site density, we still consider applying a correction to better fit the experiments and assess the validity of the other laws if provided with a proper NSD. \blue{As discussed in Subsection \ref{subsec:NSD}, this correction has to be manually done since usual correlations do not include yet surface roughness influence on the NSD,} \greenbbis{and is empirically chosen to better reproduce the measurements on those cases, taking the following form:}

\greenbbis{
\begin{equation}
    N_{sit,Li,corr} = N_{sit,Li} \times \Delta T_w^{2-0.3\Delta T_w^{0.5}}
\end{equation}
}

\greenbbis{However, this correction is only applied in the current Subsection to produce results of Figures \ref{fig:full_koss_G500} and \ref{fig:full_koss_G2000}. Its formulation is not generic and is only used here to better extract the possibility of separate effect validation using the insightful detailed experimental results of Kossolapov \cite{kossolapov_experimental_2021}. By reducing the uncertainty over the reproduction of NSD values, it allows to focus on other boiling parameters while avoiding error compensation with a possible discrepancy on the NSD.} A comparison to the traditional Kurul \& Podowski model is also presented \cite{kurul_multidimensional_1990}.

The HFP model appears to be able to reproduce correctly most of the measured quantities with a good agreement on the wall temperature, \green{bubble departure frequency}, wait time and quenching time. The latter is notably important since it means that for long wait times as in this case, estimating the quenching time $t_q$ as $t^*$ appears to be pertinent and probably is, to the best of our knowledge, the first time this hypothesis is directly verified by experimental data.

In addition, we can see that the choice of the empirical parameters for the two cases are in concordance with the evolution in the liquid mass flux: the $2000\ \debm$ case has a larger contact angle half-hysteresis $\dtheta$ and a smaller growth constant $K$. Moreover, we can note that the $500\ \debm$ case displays a small but visible static coalescence heat flux, which is consistent with having larger bubbles before departure, sliding over smaller distances due to the reduced liquid velocity. Also, the enhanced sliding in the fastest liquid case logically induces a much larger quenching heat flux.

On the contrary, we can observe that the average bubble growth time is overestimated though having a coherent evolution with the wall superheat compared to the measurements. Several factors can explain this result:

\begin{itemize}
    \item First, the hypothesis of bubble sliding until coalescing with an other one will overestimate its lifetime, especially at lower wall temperatures \ie lower nucleation site density. In this case, the average distance between two nucleating bubbles can become very large thus leading to potentially nonphysical sliding length. This is in accordance with the larger overestimation in $t_g$ being observed at low $\Delta T_w$ while a way better match is achieved at high $\Delta T_w$. On this point, detailed measurements of the average sliding length would be very valuable.

    \item Second, we do not have access to bubble departure diameter measurements in this case. Therefore, if the model overestimates it, so will be the bubble growth time. Kossolapov \cite{kossolapov_experimental_2021} also explains that nearly instantaneous departure of nucleating bubbles is observed in several experiments which could also add up to this $t_g$ overestimation.
\end{itemize}

More importantly, we note that Kurul \& Podowski original model is able to perform correctly to predict the boiling curve. However, the detailed modeling clearly shows its physical inconsistency on many aspects: 

\begin{itemize}
    \item It fails to predict the nucleation site density by two orders of magnitude ;
    \item The predicted \green{bubble departure frequency} (and wait time) are inconsistent with the experimental trend, showing very large overestimation at low superheat and reaching reasonable values at high superheat by decreasing while experiments show an increasing trend, \green{which is likely a consequence of Cole's formula initially developed for pool boiling conditions};
    \item It applies the wait time as a quenching time in the quenching heat flux, which overestimates the transient heat transfer duration by two orders of magnitude ;
    \item The predicted bubble lift-off radius is of the order of the distance between two bubbles on the wall, meaning bubbles having diameters larger than 1~mm which is inconsistent with experiments at such moderate pressures \cite{kossolapov_experimental_2021} ;
    \item The boiling curves for both cases show a very limited quenching heat flux, while sliding in vertical flow boiling is expected to largely enhance the transient heat transfer \cite{estrada-perez_time-resolved_2018, kossolapov_experimental_2021}.
\end{itemize}

Here we face a significant problem mentioned in the introduction of this work: a given HFP model can provide acceptable predictions of the couple ($\phi_{w}$, $\Delta T_w$) while largely failing to capture the physics of several boiling parameters. \bluebis{The goal of the comparison conducted here is to emphasize this issue and highlight the fact that if one hopes to reach a much more consistent modeling of wall boiling, a thorough validation of the various closure laws must be achieved beforehand.} 

Detailed measurements of the contribution of the different heat fluxes would be an insightful mean to complete the validation of the proposed modeling. Indeed, the absence of sliding in Kurul \& Podowski model finds an \green{evaporation heat flux} up to more than 90\% of the total heat flux while the present model remains around 20\% for the case of Figure \ref{fig:full_koss_G2000}. This \green{evaporation flux}, if used as an input in a CFD simulation, could then be erroneous even though proper wall temperature and heat flux are found. 

\bluebis{Moreover, we can see that both model are not completely able to precisely reproduce the entirety of the wall heat flux measurements / wall temperature measurements in the boiling region (Fig. \ref{fig:full_koss_G500} and \ref{fig:full_koss_G2000}), which may also indicate that at high heat fluxes some hypotheses such as single bubble dynamics may become erroneous, being replaced by more important bubble interactions and possibly dry patches formation when approaching to the Critical Heat Flux.}

For the sake of further exploring the validation of such models, we will directly apply them to wall heat flux predictions on a larger dataset in the next subsection.

\subsection{Wall heat flux predictions}
\label{subsec:phiw_pred}

In order to cover a larger range of pressure, we gather measurements from Kossolapov \cite{kommajosyula_development_2020, kossolapov_experimental_2021} and Jens \& Lottes \cite{jens_analysis_1951}, overall ranging from 1 bar to 138 bar. Operation conditions of these experiments are gathered on Table \ref{tab:exp_data_phiw}  

\begin{table}[h!]

%\begin{changemargin}{-1cm}{0cm}

\noindent\makebox[\textwidth]{

\scriptsize
\centering
\begin{tabular}{p{20mm}|c c c c c c c c} 
Author & $D_{h}$ [mm] & $P$ [bar] & $G_{L}$ [$\debm$] & $\Delta T_{L}$ [K] & $\phi_{w}$ [MW/m\up{2}] & $\Delta T_{w}$ [K] &$N_{mes}$ [-] \\
\hline
\\

Kossolapov \cite{kossolapov_experimental_2021} \newline (2021) & 11.78 & 1.12 - 75.8 & 500 - 2000 & 10 & 0.23 - 7.93  & 1.55 - 30.77 & 81 \\

Jens-Lottes \cite{jens_analysis_1951} \newline (1951) & 5.74 & 137.9 & 2617.5 & 53.3 - 92.2 & 2.15 - 3.63 & 1.81 - 4.16 & 38 \\

\hline
\end{tabular}
}
\caption{Experimental data range of wall temperature / heat flux measurements used for validation.}
\label{tab:exp_data_phiw}
\end{table}

Since the proposed model requires the input of the contact angle $\theta$, the contact angle half-hysteresis $\dtheta$ and the bubble growth constant $K$, we detail the values used for the validation. The contact angle value is kept constant for a given experiment with $\theta = 85\degree$ for Kossolapov cases, as the heater used in the experiment is a smooth prototypical layer of ITO \cite{kossolapov_experimental_2021}. For Jens \& Lottes experiment, precise evaluation of the contact angle in those conditions is \greenbis{complicated} but, according to Song \& Fan work \greenbis{who compiled a large amount of experimental contact angle measurements}, the water contact angle on metallic surfaces (\eg\ stainless steel) at large pressure present a strong decrease when reaching surface temperatures above $200\degree$C \greenbis{before stabilizing between $30\degree$ and $40\degree$}. Since saturation temperature of water at 138 bar is $T_{sat} \approx 335\degree$C, we then use \greenbis{$\theta =35\degree$} \cite{song_temperature_2021}.

Regarding the contact angle half-hysteresis, we consider $\dtheta = 1\degree$ for pressures above 5 bar, since bubbles will be smaller and closer to a spherical shape facing smaller liquid velocities. For pressures below 5 bar, we take $\dtheta = 3\degree$ to represent the larger tilt exerted on larger bubbles under larger liquid velocities.

Finally, values for the growth constant $K$ were adjusted depending on the cases to provide a best fitting and are summed up in Table \ref{tab:K_values}.

\begin{table}[H]
\centering
\small
    \begin{tabular}[b]{c|c} 
    Case & $K$ [-] \\
    \hline
    \\
    $P=137.9$ bar & \greenbis{1.5} \\
    \\
    $P=75.8$ bar & 1.5 \\
    \\
    $P=19.8$ bar & 1.2 \\
    \\
    $P=10.5$ bar & \makecell{\greenbis{0.7} if $G_L < 2000~\debm$ \\ \greenbis{0.5} if $G_L = 2000~\debm$ } \\
    \\
    $P=3.0$ bar & \greenbis{0.3} \\
    \\
    $P=1.1$ bar & \greenbis{1.1} \\
    \\
    \hline
    \end{tabular}
    \caption{Growth constant values}
\label{tab:K_values}
\end{table}

Though we have to acknowledge that the model was sensitive to the value of $K$, we can noteworthy observe that these best fitting values present a coherent evolution with the physical conditions:

\begin{itemize}
    \item The values of $K$ remain close to unity and below $2$, which is consistent with experimental observations for sliding bubbles at low and high pressures and theoretical models \cite{favre_updated_2023, scriven_dynamics_1959, forster_growth_1954} ;
    \item $K$ coherently increases with the pressure since bubbles will drastically diminish in size, thus staying closer to the heated wall within the superheated liquid layer and approaching the theoretical case of the bubble in an uniform temperature field ;
    \item Lower $K$ are observed for larger bubbles at low pressure where they will sooner be in contact with subcooled liquid, reducing their growth rate. Similarly, larger liquid velocities for the $10.5$ bar case of Kossolapov will increase the interfacial Nusselt number responsible for condensation \cite{ranz_evaporation_1952} ;
    \item Singular larger value of $K$ for the 1.1 bar case of Kossolapov, which could be explained by the existence of the liquid microlayer in the boiling process in those conditions that rapidly disappears when pressure overcomes 3 bar \cite{kossolapov_experimental_2021}.
\end{itemize}

In the end, this indicates that $K$ remains a proper physical parameter of the model rather than a pure numerical tuning coefficient. A dedicated modeling to estimate its value in any flow conditions would be of interest in order to exclude its manual choice, but this would require a large data-set of bubble growth profiles over broad range of physical conditions.

\begin{figure}[H]
\includegraphics[width=1.0\linewidth]{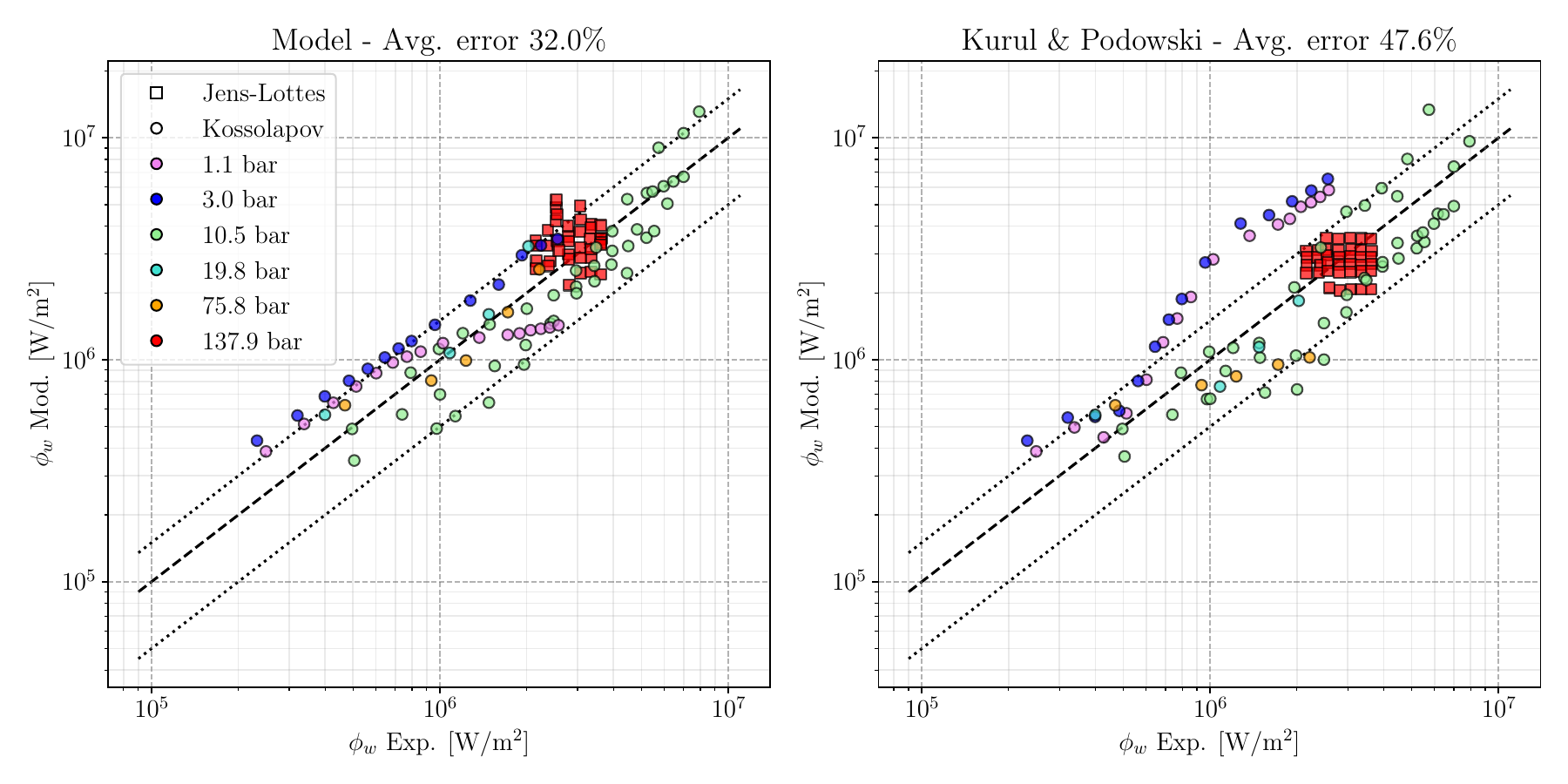}
\caption{Wall heat flux predictions on Kossolapov et Jens \& Lottes data - Proposed model and Kurul \& Podowski model. Dashed lines represent $\pm$ 50\% error bars.}
\label{fig:phiw_pred}
\end{figure}

Figure \ref{fig:phiw_pred} gathers the results of predicted total heat fluxes versus experimental results for the proposed model and Kurul \& Podowski formulation.

Overall, the new HFP formulation performs better than Kurul \& Podowski model particularly for high pressure cases with an average error below \greenbis{40\%} for the whole database. This error could be expected since it remains of the order of most of the selected closure laws accuracy. On the other hand, Kurul \& Podowksi model has an average error of \greenbis{47.6\%} which, in the end, remains not significantly worse than the proposed model. \bluebis{Additional validation on low pressure cases using Kennel \cite{kennel_local_1949} data is presented in \ref{app:kennel} and exhibits similar conclusions with the proposed model performing significantly better than Kurul \& Podowski one.} As previously mentioned, the good results produced by Kurul \& Podowski model are likely to rely on very large error compensation from closure laws that exhibit unphysical trends. Therefore, achieving a globally better result after the proposed detailed validation and assessment process is an encouraging feature supporting the need of pursuing efforts in that direction to further improve the HFP modeling.

\section{Conclusion}
\label{sec:ccl}

\bluebis{The goal of this work was to follow the whole process of building a new Heat Flux Partitioning model dedicated to vertical boiling flows while discussing the relevance of this approach in the light of recent experimental results from the literature. The main idea was to thoroughly focus on closure laws validation to enforce their physical consistency} with the hope of ending up with an overall better modeling compared to traditional approaches.

The main features regarding model development are:

\begin{itemize}
\item The use of a detailed force balance model for bubble departure and sliding, previously validated in details for a large range of operation conditions \cite{favre_updated_2023}.

\item Development of a maximum bubble lift-off diameter correlation based on 212 experimental measurement from the literature, with a mathematical formulation ensuring its applicability for saturated or quiescent flows.

\item Comprehensive selection of closure laws avoiding strong physical assumptions (\eg bubble growth time negligible versus the wait time) with detailed validation using existing measurements of bubble wait time and nucleation site density.

\item Accounting for bubble interactions by assuming a coalescence induced bubble lift-off and a statistical approach for nucleation site suppression and static coalescence between non-sliding nucleating bubbles.

\item Reduced number of remaining arbitrary parameters (contact angle, contact angle hysteresis, bubble growth constant) chosen consistently considering the physical operation conditions of simulated cases.  

\end{itemize}

Then, the validation process highlighted the following points:

\begin{itemize}
\item Cross-comparison of both the boiling curves and boiling sub-parameters such as nucleation site density and nucleation time scales can be all globally satisfying thanks to the previous detailed assessment of closure laws. 

\item The model is able to capture different heat flux partitioning regime, \eg\ transition from situations where \green{evaporation heat flux} is dominant to quenching dominant regimes when high liquid velocity increases the magnitude of bubble sliding.

\item Older models like that of Kurul \& Podowski \cite{kurul_multidimensional_1990} present incoherent trends when compared to experiments for several parameters such as the \green{bubble departure frequency} or the wait time. However, they lead to error compensation resulting in a wall heat flux / temperature prediction that could seem acceptable at first glance, which should be a warning to avoid general application of such models without proper validation beforehand.

\item Global predictions of wall heat flux values can be achieved with a slightly better accuracy than with traditional models, along with the use of physically consistent values for the contact angle, angle hysteresis and bubble growth constant. Which is a satisfying feature meaning that the model construction has not led to the need of nonphysical tuning of these parameters.

\end{itemize}

\bluebis{All in all, the main conclusion of the present work is the highlight of the importance of conducting a detailed validation process throughout the development of future Heat Flux Partitioning Model.} Indeed, the present approach demonstrated that relying on simpler assessment such as sole total heat flux predictions was not enough and left room for significant error compensation from the closure laws, thus resulting in an erroneous estimation of the different heat transfer mechanisms, \ie failing to predict the partitioning between sensible and latent heat transfer though providing an overall acceptable wall temperature / total heat flux couple. In addition, if included in a CFD framework, such discrepancies would results in feeding the simulation with wrong sensible and latent heat transfer as boundary conditions at the wall, resulting in errors on the \green{evaporation heat flux} and bubble lift-off diameter which would negatively impact the predictions of the vapor phase dynamics, \greenbbis{at least close to the wall}.

In the light of this observation, one may more generally question the pertinence of the Heat Flux Partitioning approach for wall boiling modeling. Without a proper and comprehensive selection of the closure laws, simpler direct correlations similar to older ones \cite{thom_boiling_1967, jens_analysis_1951} \bluebis{or more direct modeling switching the whole heat flux towards boiling after the Onset of Significant Void \cite{reiss_heat_flux}} may be more precise if tailored for the desired configuration. The goal of the Heat Flux Partitioning is to provide the wall boiling modeling with a deeper and more generic physical description of the phase change phenomenon and must then be assembled accordingly by avoiding correlations with narrow ranges of application and rely on consistent values for any remaining empirical parameters for its full closure. 

Fortunately enough, an increasing number of experimental works are now providing insightful measurements for the development of future HFP models \cite{kossolapov_experimental_2021} and are shedding light on new physical mechanisms at stake in the wall boiling phenomenon. This may help to further leverage some limitations of the current modeling \eg\ by helping the development of a more general modeling of the bubble growth or give better evaluations of the growth constant $K$. One can also mention the work of Song \& Fan \cite{song_temperature_2021} which is of great interest to tackle the contact angle value problem, \blue{which value should theoretically be evolving with the wall temperature / heat flux and flow conditions. The perspective of accounting for such a behavior would surely be a significant step towards a more physical modeling of wall boiling, though it will require extensive measurements over surfaces, fluids an physical conditions of interest.} \green{Finally, the various bubble behaviors could be further investigated through wall bubble tracking methods \cite{kossolapov_bubble_2024} in order to estimate the nucleation site densities leading to bubble sliding and / or coalescence thus allowing to include a more realistic categories of bubble interactions and dynamics on the boiling surface.}

\greenbbis{In its current state, the model showed a limitation regarding its CFD implementation due to a significant increase in the overall computation time related to the time-iterative loop for bubble sliding computation (low superheat or liquid mass flux leading to slow sliding over long distances). However, one may expect that bulk void fraction profiles in subcooled boiling flows will exhibit small sensitivity to the change in modeling, as bulk heat and mass transfer models often overcome the effect of changes in HFP model \cite{favre_modelisation_2023, gilman_self-consistent_2017}. Nonetheless, a specific work to optimize the bubble sliding calculation and achieve a complete validation of the CFD implementation of the model is currently ongoing \cite{fayet_modelisation_2025}.}

\section{Acknowledgements}

This work was funded by \'Electricit\'e de France (EDF) in collaboration with Institut de M\'ecanique des Fluides de Toulouse (IMFT).

\appendix

\section{\bluebis{Additional wall heat flux predictions at low pressure}}
\label{app:kennel}

%\greenbis{
This first appendix serves as an extra validation database complementary to Subsection \ref{subsec:phiw_pred}. Here, we use the experimental results of Kennel \cite{kennel_local_1949} who conducted wall temperature measurements in a boiling upwards annular flow with heated stainless steel. Table \ref{tab:exp_data_kennel} sums up the operating conditions corresponding to Kennel's measurements database. 

\begin{table}[H]
\noindent\makebox[\textwidth]{

\scriptsize
\centering
\begin{tabular}{p{20mm}|c c c c c c c c} 
Author & $D_{h}$ [mm] & $P$ [bar] & $G_{L}$ [$\debm$] & $\Delta T_{L}$ [K] & $\phi_{w}$ [MW/m\up{2}] & $\Delta T_{w}$ [K] &$N_{mes}$ [-] \\
\hline
\\

Kennel \cite{kennel_local_1949} \newline (1949) & 4.3 - 13.2 & 2.1 - 6.2 & 284 - 10\ 577 & 11.1-83.3 & 0.053 - 6.35  & 1.64 - 49.7 & 224 \\

\hline
\end{tabular}
}
\caption{Experimental data range of wall temperature / heat flux measurements used for Kennel database.}
\label{tab:exp_data_kennel}
\end{table}

For this dataset, the contact angle is set to $\theta = 70\degree$ which corresponds to a usual value for water on stainless steel at temperatures around 100°C \cite{song_temperature_2021}. The contact angle half-hysteresis has been set to $\dtheta = 3 \degree$ in accordance with Subsection \ref{subsec:phiw_pred} where a larger angle hysteresis is expected at low pressures inducing larger bubbles. Regarding the value of the bubble growth constant $K$ was adjusted depending on the cases and its values are summed up in Table \ref{tab:K_Kennel}.

\begin{table}[H]
\centering
\small
    \begin{tabular}[b]{c||c c c c} 
     ~ & $\Delta T_L = 11$°C & $\Delta T_L = 27$°C &$\Delta T_L = 55$°C & $\Delta T_L = 83$°C\\
    \hline
    \hline
    \\
    $P < 3$ bar & 1.5 & 1.2. & 1.0 & 0.5 \\
    \\
    $P > 3$ bar & 0.2 & 0.2 & 0.2 & 0.1 \\
    \\
    \end{tabular}
    \caption{Bubble growth constant values $K$ [-] for Kennel cases}
\label{tab:K_Kennel}
\end{table}

We note that the values of $K$ that were found to be great to match Kennel data are exhibiting trends that are coherent with those in Table \ref{tab:K_values}:

\begin{itemize}
    \item For cases at pressures lower than 3 bar, larger values of $K$ are needed, also possibly pointing at the impact of microlayer formation accelerating the bubble growth and vanishing if $P>3$ bar \cite{kossolapov_experimental_2021}.
    \item The overall trend is a decrease of $K$ with the liquid subcooling $\Delta T_L$, which is coherent since large subcooling induces slower bubble growth.
    \item At pressures above 3 bar, quite low values of $K$ are being used, indicating that the large subcooling used in Kennel cases may strongly be impacting bubble growth. These value seem to be coherent with the Kossolapov case at 3.0 bar (Table \ref{tab:K_values}).
\end{itemize}

\begin{figure}[H]
\includegraphics[width=1.0\linewidth]{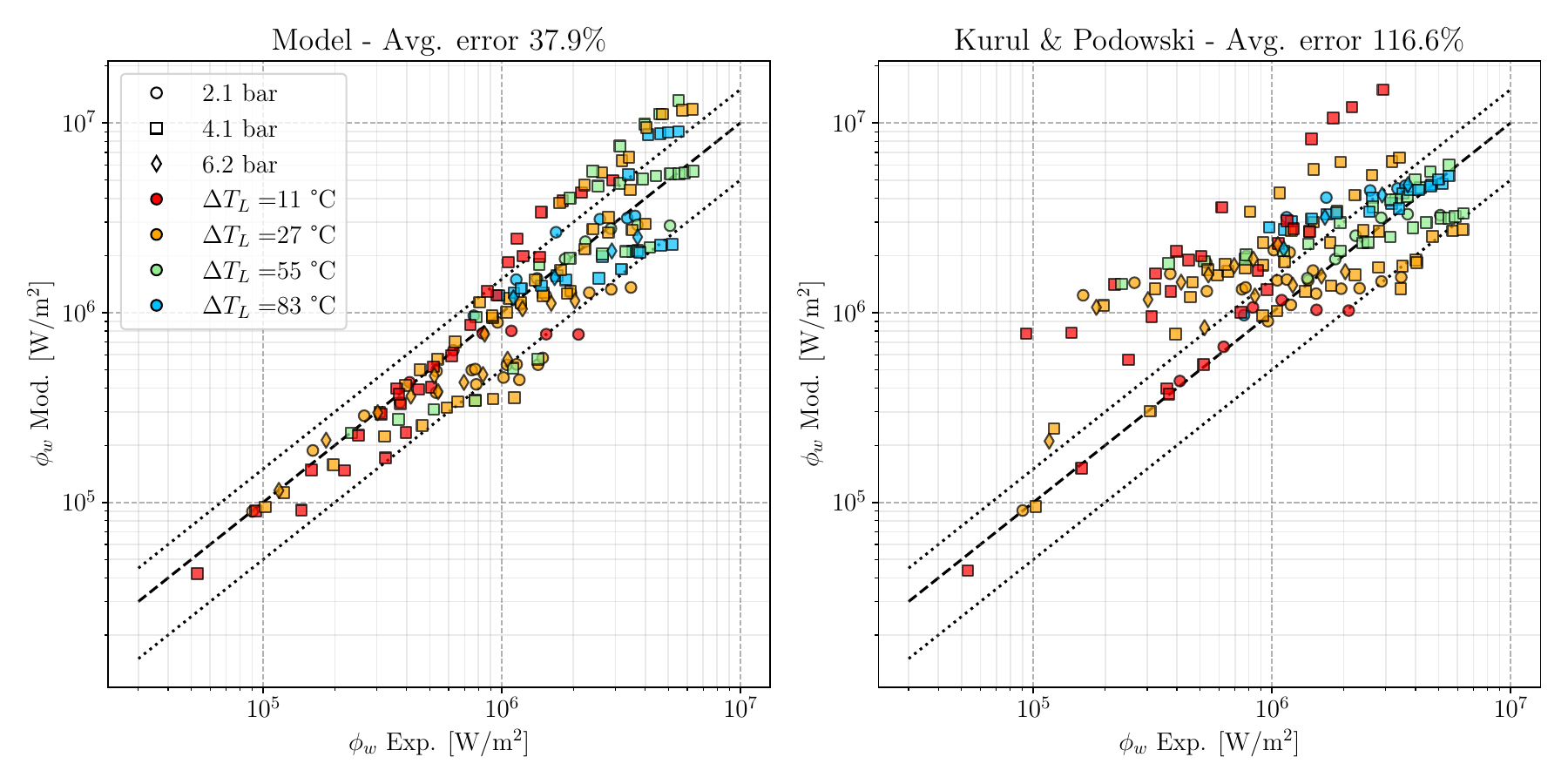}
\caption{Wall heat flux predictions on Kennel data - Proposed model and Kurul \& Podowski model. Dashed lines represent $\pm$ 50\% error bars.}
\label{fig:phiw_Kennel}

We note that the proposed model is able to perform better than the one of Kurul \& Podowksi by relying on a physically consistent selection of the closure parameters $\theta$, $\dtheta$ and $K$. Moreover, the values of $\dtheta$ and $K$ are in accordance with other datasets explored in Subsection \ref{subsec:phiw_pred}, yielding an overall accuracy lower than 40\%, similar to the result for Kossolapov and Jens-Lottes data (Figure \ref{fig:phiw_pred}). This further insists on the physical meaning of those parameters that would ultimately benefit of having their own modeling if more dedicated experimental data would be available.
\end{figure}

Comparison between the proposed model and Kurul \& Podowski approach on Kennel data is presented on Figure \ref{fig:phiw_Kennel}.

%}

\section{\greenbbis{Heat Flux Partitioning sensitivity to $\theta$, $\dtheta$ and $K$}}

\begin{figure}[H]
\centering

\subfloat[$\theta - 10\%$]{
\includegraphics[width=0.49\linewidth]{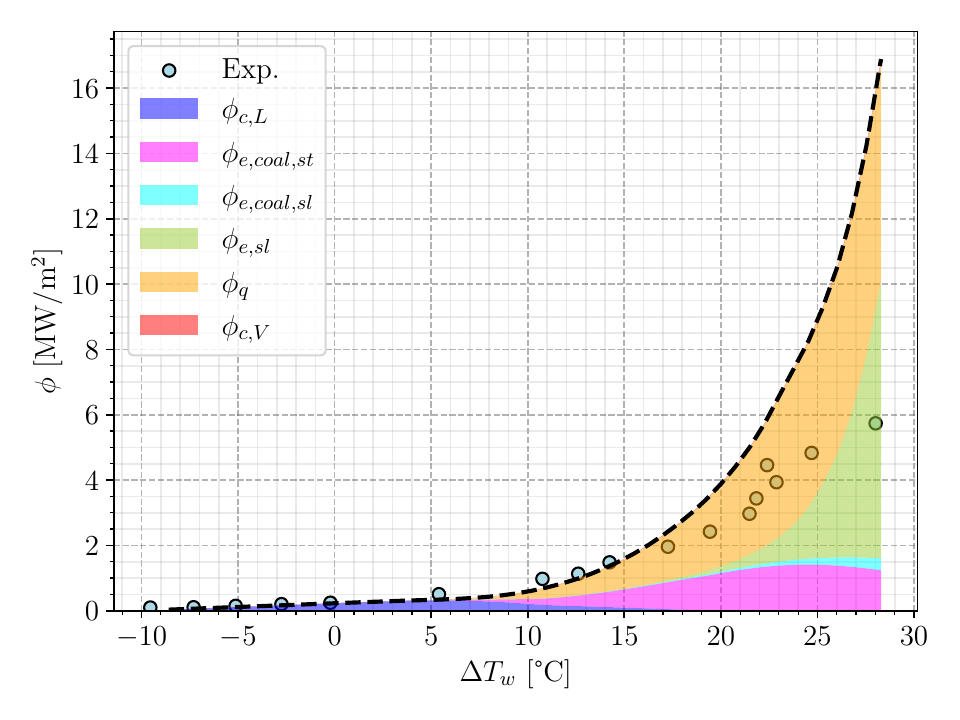}
}
\subfloat[$\theta + 10\%$]{
\includegraphics[width=0.49\linewidth]{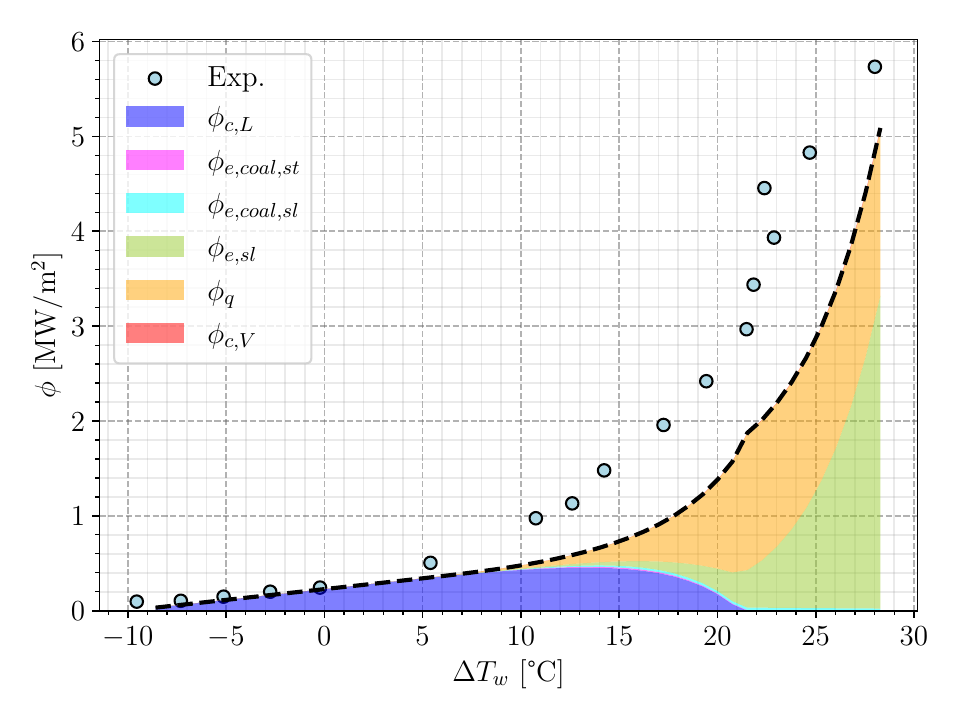}
}

\subfloat[$\dtheta - 10\%$]{
\includegraphics[width=0.49\linewidth]{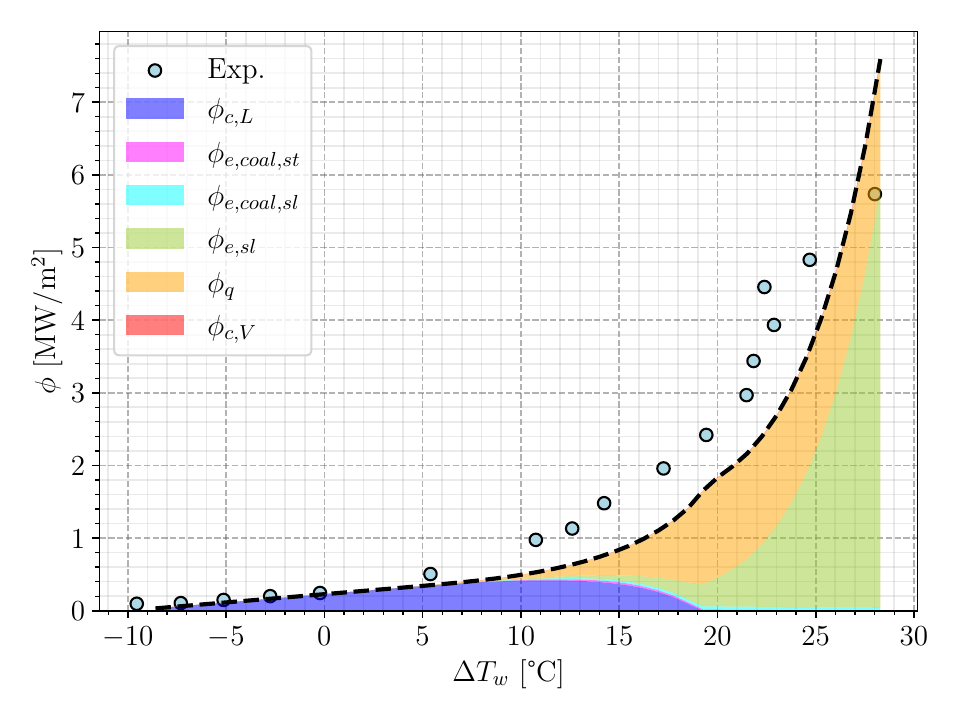}
}
\subfloat[$\dtheta + 10\%$]{
\includegraphics[width=0.49\linewidth]{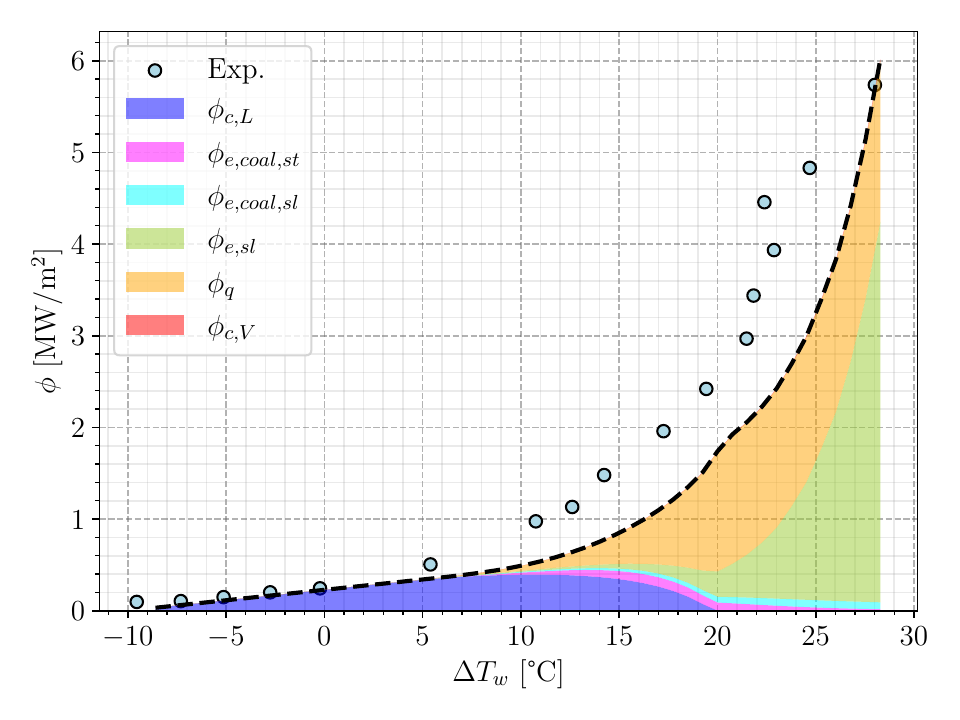}
}

\subfloat[$K - 10\%$]{
\includegraphics[width=0.49\linewidth]{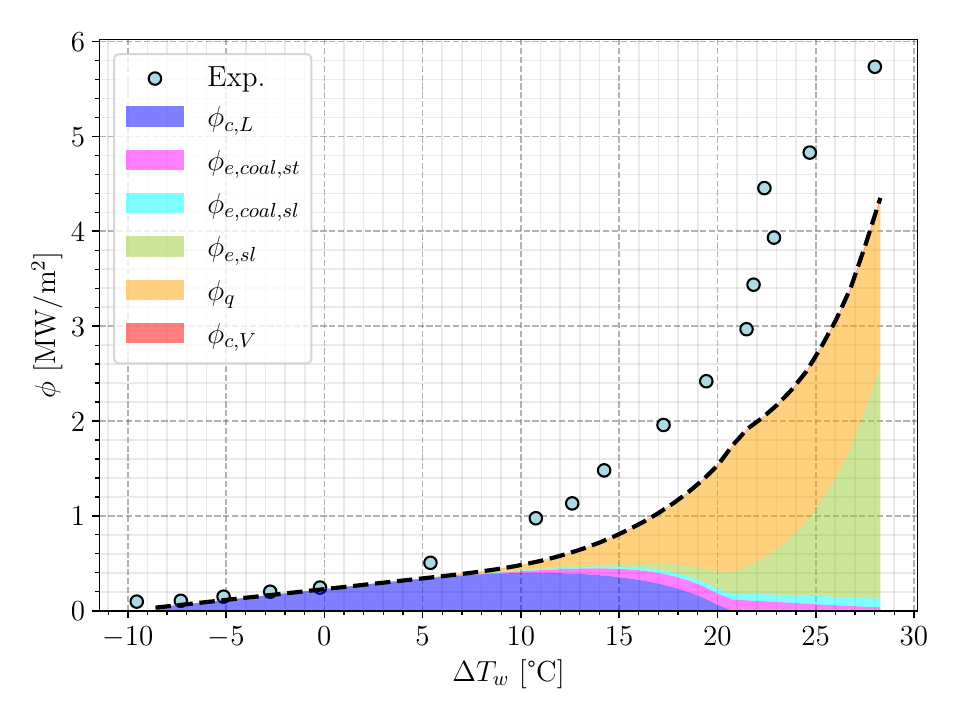}
}
\subfloat[$K + 10\%$]{
\includegraphics[width=0.49\linewidth]{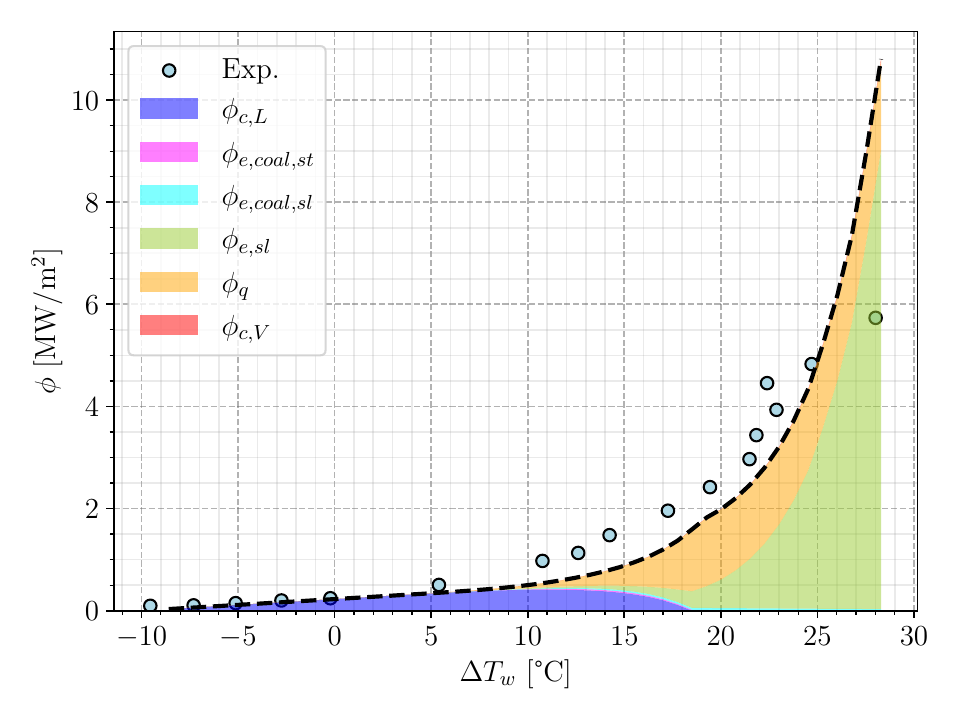}
}

\caption{Sensitivity of the model to $\pm 10\%$ changes on the empirical parameters.}
\label{fig:sensi_analysis}
\end{figure}

Since the final model has been reduced to an empiricism laying over three parameters, namely the contact angle $\theta$, the contact angle half-hysteresis $\dtheta$ and the bubble growth constant $K$, this Appendix aims at showing some sensitivity analysis of the  model to change in those parameters' values. The reference case taken here is the Kossolapov case at $G = 500 \debm$ from Figure \ref{fig:full_koss_G500} ($\theta=85\degree$, $\dtheta=2\degree$, $K=0.7$), for which we plot the boiling curve under changes of $\pm$ 10\% on $\theta$, $\dtheta$ and K. Results are presented on Figure \ref{fig:sensi_analysis}.

The model seems to be quite sensitive to the value of the contact angle $\theta$, which makes sense since it influences a large number of closure laws, among which the NSD, the wait time and the capillary force in the bubble force balance. The combined change of all those parameters at the same time under the change of $\theta$ naturally impacts significantly the resulting boiling curve and heat flux partitioning. Especially for this case, reducing the value of $\theta$ decreases slightly the NSD but also largely decreases the wait time $t_w$, resulting in a higher bubble departure frequency $f$ and thus in a larger static coalescence heat flux $\phi_{e,coal,st}$. The effect of increasing $\theta$ is the exact opposite, suppressing the remaining static coalescence heat flux.

On the other hand, the effect of a change in $\dtheta$ seems quite limited since its only influence is in the bubble force balance, controlling the bubble departure diameter through the magnitude of capillary force (Eq. \ref{eq:pred_nogr}). Increasing $\dtheta$ increases the bubble departure diameter by sliding, thus increasing the probability of bubble coalescence on the wall leading to larger coalescence fluxes $\phi_{e,coal,st}$ and $\phi_{e,coal,sl}$.

Finally, the model also shows a significant sensitivity to the value of $K$ since the latter will impact the bubble force balance along with the bubble growth time, thus impacting the overall bubble dynamics and nucleation cycle time-scales. A lower value of $K$ naturally leading to slower growing bubbles staying longer on their nucleation sites and thus enhancing the probability of coalescence, also increasing $\phi_{e,coal,st}$ and $\phi_{e,coal,sl}$. On the other hand, increasing $K$ results in overall larger bubbles on the boiling surface, naturally increasing the evaporation heat flux associated to non-coalescing bubbles $\phi_{e,sl}$

All in all, we have to acknowledge that the efforts to reduce the empiricism of the model as much as possible logically concentrates all its sensitivity in the remaining empirical parameters, here being mainly $\theta$ and $K$. However, the best predictions obtained in the frame of this work were at least satisfyingly produced using physically coherent values for those parameters, highlighting the need for detailed experiments to enable the development of dedicated models and even further reduce the remaining empiricism laying upon HFP models.

\begin{figure}[H]
\centering

\includegraphics[width=0.49\linewidth]{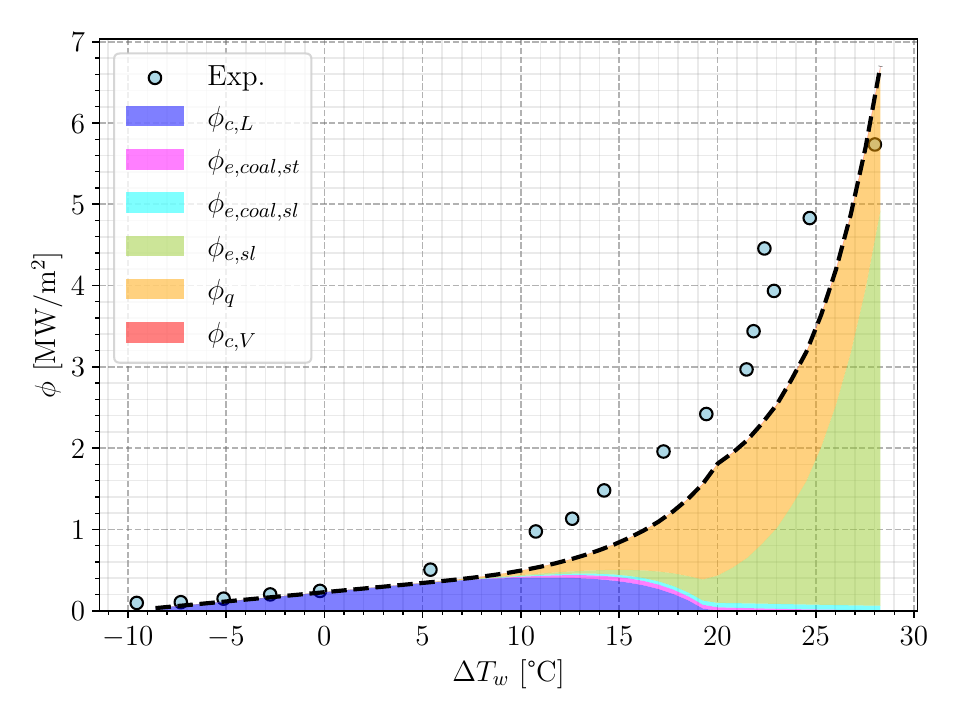}
\caption{Reference case (Figure \ref{fig:full_koss_G500})}
\end{figure}

\section{Heat Flux Partitioning algorithm and usage details}

\begin{figure}
    \centering
    \includegraphics[width=1.0\linewidth]{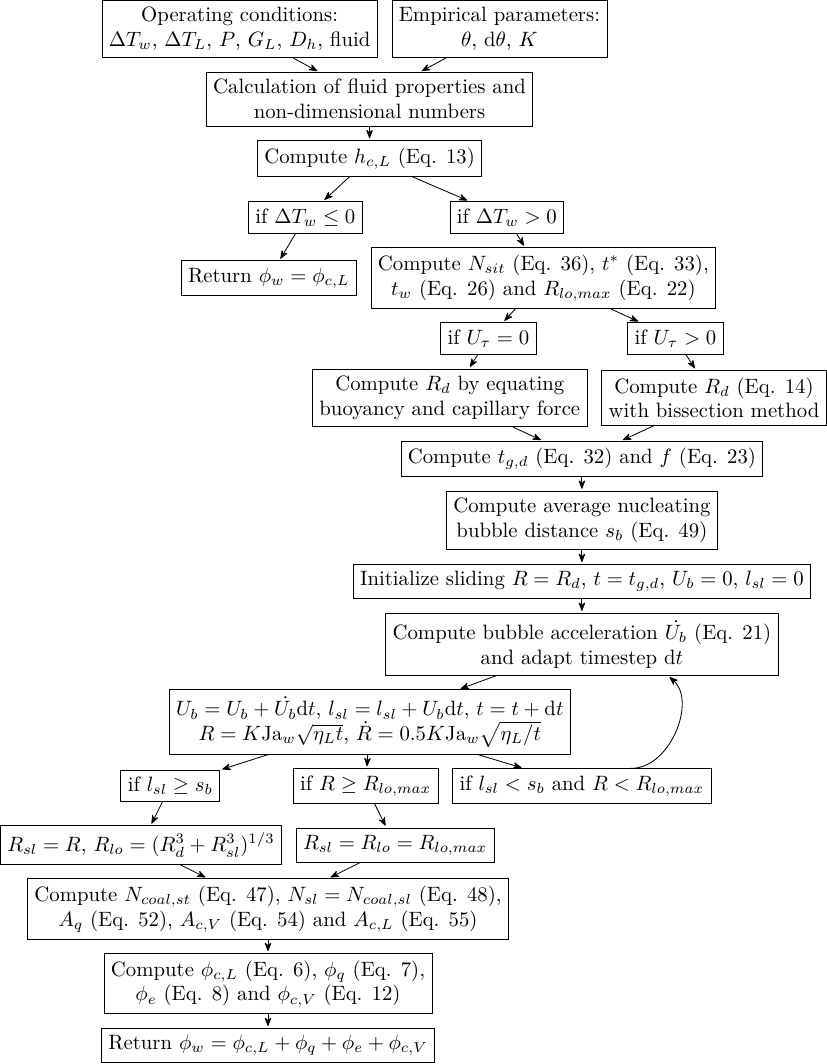}
    \caption{Sketch of the algorithm process for HFP calculation}
    \label{fig:algo_hfp}
\end{figure}

The code containing the implementation of the HFP model in python and used to produce the results of Section \ref{sec:validation} is available at \url{https://github.com/lucfavre/hfp_model}. The fluid properties have been calculated using the CoolProp package \cite{bell_pure_2014}. A sketch of the algorithm process followed to perform the whole heat flux partitioning calculation is presented on Figure \ref{fig:algo_hfp} in the case where the wall superheat $\Delta T_w$ is known. If the total heat flux $\phi_w$ is taken as an input, an iterative process over $\Delta T_w$ values has to be used (\eg bisection method) to find the value of $\Delta T_w$ giving $\mathrm{HFP}\parth{\Delta T_w} = \phi_w$.

Finally, we remind here a few guidelines to select physically-consistent values for the remaining empirical parameters:
\begin{itemize}
    \item Contact angle $\theta$:
    \begin{itemize}
        \item Water on metallic prototypical surfaces: From $90\degree$ at low temperature down to lower than 20\degree\ at high pressure and temperature \cite{song_temperature_2021} ;
        \item Water on non-prototypical smooth surfaces: Close to 80\degree\  
        \cite{kossolapov_experimental_2021} ;
        \item Refrigerants on metallic surfaces: Down to lower than $10\degree$ \cite{zou_heating_2013}.
    \end{itemize}
    \item Contact angle half-hysteresis $\dtheta$:
    \begin{itemize}
        \item Up to more than $20\degree$ for large bubbles (\eg low pressure, high heat fluxes) ;
        \item Down to lower than $1\degree$ for very small bubbles (\eg high pressure, low heat fluxes).
    \end{itemize}
    \item Bubble growth constant $K$:
    \begin{itemize}
        \item Between 0.1 and 1 for large liquid subcooling, high liquid velocities and low pressure ;
        \item Between 1 and 2 for moderate to low liquid subcooling, approaching 2 for high pressure flow (\ie very small bubbles).
        \item \greenbbis{Above 1 for water at pressures lower than 3 bar (possibility of microlayer appearance enhancing bubble growth)}
    \end{itemize}
\end{itemize}

\newpage
%\section{References}
%\bibliographystyle{plain}
\bibliographystyle{elsarticle-num-names}
\bibliography{bib_2}

\end{document}